\documentclass[prb,twocolumn,showpacs,preprintnumbers,amsmath,amssymb,floatfix]{revtex4}
\usepackage{amssymb}
\usepackage{amsmath}
\usepackage{dsfont}
\usepackage{graphicx}
\usepackage{dcolumn}
\usepackage{bm}
\newcommand{\PRL}[3]{Phys. Rev. Lett. {\textbf{#1}, {#2} (#3)}}
\newcommand{\PRB}[3]{Phys. Rev. B {\textbf{#1}, {#2} (#3)}}
\newcommand{\PR}[3]{Phys. Rev. {\textbf{#1}, {#2} (#3)}}
\newcommand{\ZPB}[3]{Z. Phys. B {\textbf{#1}, {#2} (#3)}}
\newcommand{\EurPhysJB}[3]{ Eur. Phys. J. B {\textbf{#1}, {#2} (#3)}}
\begin{document}
\title{Quantum noise in ac-driven resonant-tunneling double barrier structures: Photon-assisted tunneling vs. electron anti-bunching}
\author{Jan Hammer and Wolfgang Belzig}
\affiliation{Quantum Transport Group, Universit\"at Konstanz, 78457 Konstanz, Germany}
\date{\today}
\begin{abstract}
  We study the quantum noise of electronic current in a double barrier
  system with a single resonant level. In the framework of the
  Landauer formalism we treat the double barrier as a quantum coherent
  scattering region that can exchange photons with a coupled electric
  field, e.g.  a laser beam or a periodic ac-bias voltage. As a
  consequence of the manyfold parameters that are involved in this
  system, a complicated step-like structure arises in the
  non-symmetrized current-current auto correlation spectrum and a
  peak-like structure in the cross correlation spectrum with and
  without harmonic ac-driving.  We present an analytic solution for
  these noise spectral functions by assuming a Breit-Wigner lineshape.
  In detail we study how the correlation functions are affected by
  photo-assisted tunneling (PAT) events and discuss the underlying
  elementary events of charge transfer where we identify a new kind of
  contribution to shot-noise. This enables us to clarify the influence
  of a not centered irradiation of such a structure with light in
  terms of contributions originating from different sets of coherent
  scattering channels. Moreover we show how the noise is influenced by
  acquiring a scattering phase due to the complex reflection
  amplitudes that are crucial in the Landauer approach.
\end{abstract}
\maketitle
\section{Introduction}
As a striking consequence of charge quantisation shot noise can be
used to characterize electron transport in mesoscopic
systems~\cite{LevitovLesovik93,BlanterButtiker00,Schonenberger03,Nazarov03}.
In ballistic electron transport\cite{Wharam88,vanWees88} partitioning
of the scattered quasiparticles~\cite{Buettiker94} is the mechanism
defining the statistics of charge fluctuations in the two leads.
Indeed, the principle of counting individual charges leads to the full
counting statistics approach~\cite{LevitovLee96} which has been
successfully applied to tackle a variety of problems,
e.~g. superconducting heterostructures \cite{Belzig01a,Belzig01b}, electron
transport in multi-terminal conductors~\cite{Nazarov02},
zero-frequency noise in multi-level quantum dots~\cite{Belzig05} or
frequency-dependent noise in interacting conductors.~\cite{Emary07}
This formalism has also been incorporated to characterize the
elementary events of current-current correlations for
energy-independent scattering at zero frequency but with finite
ac-driving voltage.~\cite{VanevicBelzig07,VanevicBelzig08} Such an
harmonic voltage dependence can be induced by irradiating the
structure with light~\cite{PedersenButtiker98}, e.g a laser
beam.~\cite{Erbe06} Ongoing effort in improving the detection of
current-current correlations at high frequencies
\cite{Gabelli08,Gabelli09,ZakkaBajjani10} and coupling such structures
to light fields or ac-bias
voltages~\cite{Grifoni98,LesovikLevitov94,Brune97,Schoelkopf97,Moskalets02,Kohler05,Moskalets08}
offers an interesting playground to examine quantum charge transport
or light-matter interaction in mesoscopic systems. Within the last
years considerable progress in ac-transport has been
achieved. E.g. the irradiation induced opening of a dynamical gap has
been calculated~\cite{Fistul07} in a 2D electron gas when spin-orbit
interaction is present. The 
current and noise through long, ac-driven molecular
wires~\cite{Camalet04,Kaiser07,Safi08,Safi09}, various aspects about ac-driven
carbon based
conductors~\cite{Trauzettel07,Fistul07b,Syzranov08,Zeb08,Rocha10},
photo-assisted noise in the fractional quantum Hall
regime~\cite{Crepieux04}, low-frequency current noise in diffusive
conductors~\cite{Bagrets07}, noise in adiabatic pumping
\cite{Moskalets04,Moskalets06,Moskalets07} and the influence of
electron-phonon interaction has been studied \cite{Aguado04}.
Even more works have investigated the influence of Coulomb repulsion on the
transport through a quantum dot
\cite{Sukhorukov01,Dong04,Djuric05,Thielmann05a,Thielmann05,Braun06,Dong08}.
Electron-electron interactions can be included within a Green function formalism or generalized master-equation approach.
Interestingly, quantum noise spectra are symmetrized  by performing a Markov approximation~\cite{Engel03,EngelLoss04}.
In this classical limit, one can make use of the Mac Donald formula and calculate the noise of a quantum dot system with ac-bias voltages
up to a Born approximation as shown recently in Ref.~\cite{Wu10}.  

It has been shown recently in
experiment~\cite{Gabelli08,Gabelli09,ZakkaBajjani10,Glattli09,Reydellet03,Reulet03}
and
theoretically~\cite{Lesovik97,GavishImry00,GavishImry01,Beenakker01,Schonenberger03,Engel03,EngelLoss04,EntinWohlman07,Rothstein08,Nazarov03}
that the noise of a two-terminal device, as for a coherent scattering
double barrier structure, leads to an asymmetric noise spectrum in the
quantum regime. Since current operators at different times do not
commute one could argue that, in order to get physical results, the
shot-noise spectrum should be symmetrized in the frequency $\Omega$ in
analogy to the classical noise~\cite{PedersenButtiker98}.  Indeed,
such a quantity describes experiments in the classical detection
regime correctly \cite{Lesovik97,Bednorz08,Bayandin08,Bednorz10,Bednorz10a}.
Nevertheless it has become clear during the last 
years that asymmetric noise can be measured if a detector
discriminates between the absorption and emission of energy quanta
$\hbar \Omega$ from or to the system~\cite{GavishImry00,GavishImry01}.
Then the positive (negative) frequencies of the noise spectra
correspond to energy quanta $\hbar \Omega$ transferred from (to) the
radiation field to (from) the charge carriers in the quantum dot.  The
negative frequency part of the spectrum, the emission branch, should
be measured by an active detector setup\cite{GavishImry01}, since at
low enough temperature the energy transfer from the quasiparticles to
the radiation field is forbidden otherwise.  The detected current
fluctuations are described by a combination of the ``pure''
correlators of two currents at different times.  Fourier
transformation to the frequency domain defines the asymmetric noise
spectrum, which might in addition depend on some harmonic driving in
the leads $eV_{ac} \cos(\omega t)$, as 
\begin{equation}
  S_{\alpha \beta}(\Omega, \Omega', \omega)= \int\limits_{-\infty}^{\infty} dt dt' 
  S_{\alpha \beta}(t,t',\omega)
  e^{i \Omega t + i \Omega' t'} \, .   
  \label{noisedef}
\end{equation}
The non-symmetrized shot noise correlates currents at two times:
\begin{equation}
S_{\alpha \beta}(t,t',\omega)=\left\langle \Delta \hat{I}_{\alpha}(t) \Delta \hat{I}_{\beta}(t') \right\rangle
\end{equation}
with variance $\Delta \hat{I}_{\alpha}(t) = \hat{I}_{\alpha}(t) -
\langle\hat{I}_{\alpha}(t)\rangle$. 
Experimental accessible are the fluctuations averaged over a timescales large
compared to the one defined by the driving frequency $\omega$.  Thus,
as in Ref.~\cite{PedersenButtiker98}, we introduce Wigner coordinates
$t=T+\tau/2$ and $t'=T-\tau/2$ and average over a driving period
$2\pi/\omega$. Then the noise spectrum is defined by the quantum
statistical expectation value of the Fourier-transformed
current-operator $\hat{I}_{\alpha}(\Omega)$ via $S_{\alpha
  \beta}(\Omega, \Omega',\omega)=2\pi
\delta(\Omega+\Omega')S_{\alpha
  \beta}(\Omega,\omega) =
\langle\hat{I}_{\alpha}(\Omega)\hat{I}_{\beta}(\Omega')\rangle$.
$S_{\alpha \beta}(\Omega,\omega)$  is just the Fourier transform of $S_{\alpha \beta}(\tau,\omega)$.
Similarly, in the case without ac-driving the noise is only a function
of relative times $\tau=t-t'$. In order to keep notation short,
in the dc-bias limit we write $S_{\alpha \beta}(\Omega):=S_{\alpha
  \beta}(\Omega,\omega=0)$.

As we show in this article, the finite frequency current noise can be
interpreted by splitting it into contributions which are emanating
from reservoir $\alpha={L,R}$ and being scattered into terminal
$\beta=L,R$.  This motivates us to study the individual auto-terminal
and cross-terminal current-current correlators, the `building-blocks'
of the possible noise spectra measured in
experiments~\cite{Schonenberger03,Reulet03,Reydellet03,Gabelli08,Glattli09,Gabelli09,ZakkaBajjani10}.
The paper is organized as follows: Below we describe the basic
properties of the driven quantum dot.  In the next section we provide
the basic formulas in the scattering formalism. The main results are
discussed in the two following parts about the auto- and
cross-correlation noise spectra. We relate features of the calculated
plots to possible scattering events and compare the spontaneous PAT
events at finite dc-bias without driving with those induced by the
ac-voltage. Where possible we connect our approach to known cases.
The following part is devoted to interpret the results in terms of
elementary events of charge transfer.  The results are summarized in
the last section of this article.

For resonant tunneling with energy-dependent transmission through the
scattering region, e.g. as in many quantum dots or molecules, the
scattered particles have to be in resonance with the available energy
levels of the scatterer.  In case of a single resonant level at least
one of the chemical potentials of the reservoirs has to be aligned
with this energy level.  Alternatively a quasiparticle in the leads
has to absorb or emit suitable energy quanta to bridge the energy gap
between the chemical potential and the resonance.  This can be
achieved via absorption or emission of photons stimulated by an
external electric field, typically a microwave or laser beam. In the
Tien-Gordon theory~\cite{TienGordon62,PedersenButtiker98} such an
illumination with light corresponds to an oscillating voltage in
either one or both leads.  Depending on the way the light field is
coupled to the electronic circuit it has to be treated as either
symmetric or asymmetric in the amplitudes of the harmonic ac-driving
in the left and right lead. A spatial asymmetry in the illumination
could additionally introduce different temperatures in the two leads
and thus create thermo-currents~\cite{Moskalets02}.  If the driving is
asymmetric there can be a photocurrent even when no bias voltage is
applied.  For the noise, asymmetry effects in terms of enhancement or
reduction of the ac-drive in a terminal $\alpha$ can be related to the
corresponding correlator and so to the kind of scattering events
described by its integrand.  We neglect interactions and disregard
charging effects by assuming metallic structures with pefect
screening. In general one should treat charging effects in a
self-consistent manner via a dynamical
conductance\cite{PedersenButtiker98,ButtikerPretre93,ButtikerChristen96,ChristenButtiker96},
which has been recently confirmed experimentally \cite{Gabelli07}
\section{Scattering approach to resonant tunneling with ac-driving\label{sec:scattering}}
Following the work of Pedersen and B\"uttiker~\cite{PedersenButtiker98}
we take the ac-voltage $V_{m}(t)=V_{ac,{m}}\cos\left(\omega t \right)$ at contact $m=L,R$ 
into account by redefinition of the reservoir operators via
$\hat{a}_{m}(\epsilon) =\sum_l \hat{a}'_{m}(\epsilon - l\hbar \omega) J_l\left(\alpha_m\right)$.
\begin{figure}[t]
\begin{center}
\includegraphics[width=0.99\columnwidth]{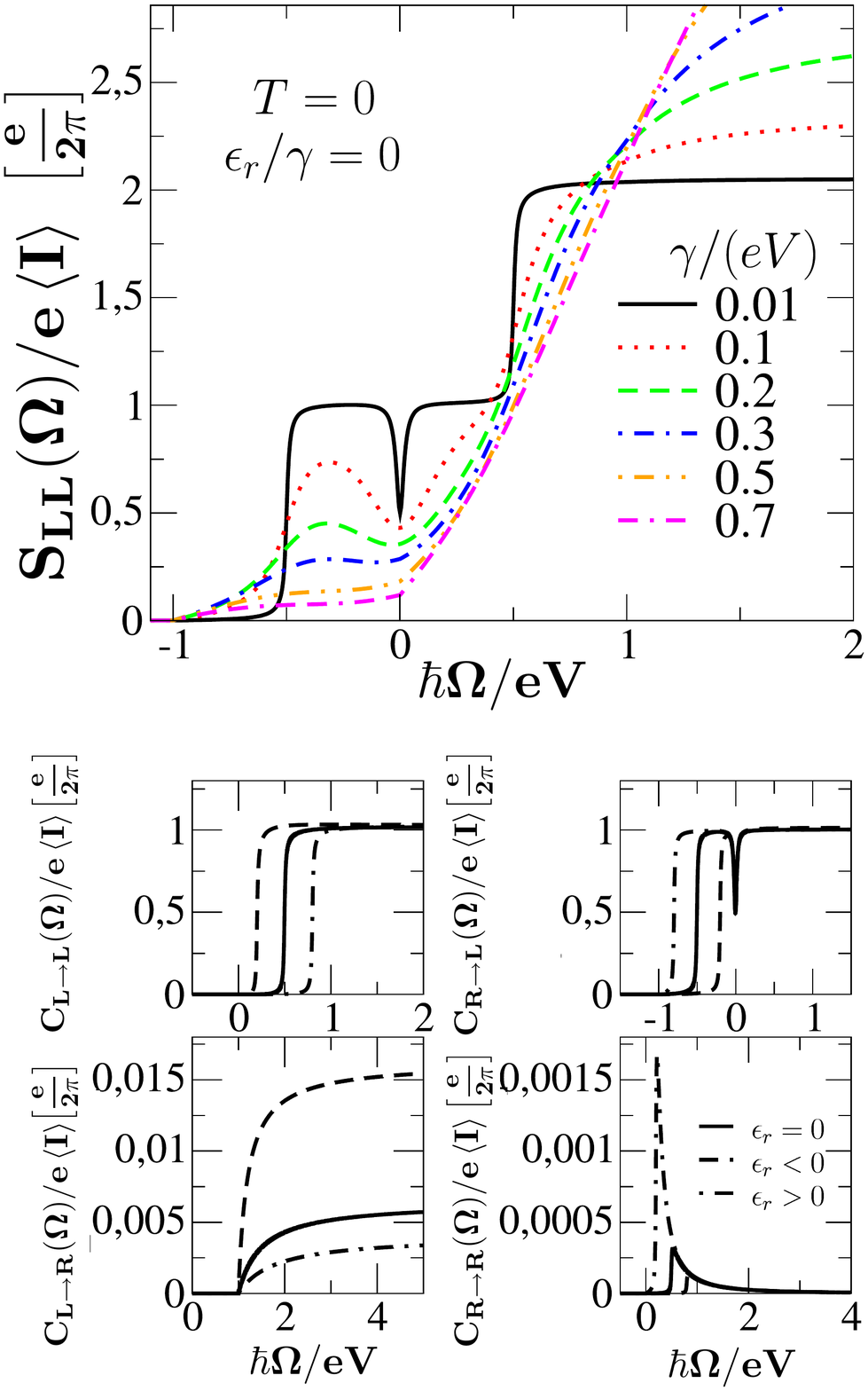}
\caption{
	\label{fig-1} Top: Zero-temperature auto-correlations for a centered resonance (top) where each curve 
	belongs to a different resonance width $\gamma$.  The curves are for symmetrically 
	applied dc-bias ($-eV/2 = \mu_l = -\mu_r $).
	With increasing resonance width $\gamma$ the step-like structure of the 
	noise spectrum gets washed out. By increasing $\gamma$ the noise 
	at negative frequencies is reduced while the spectrum 
	exhibits the typical linear frequency-dependence for energy-independent scattering.
	Bottom: The four contributions to the noise spectrum 
	$S_{LL}(\Omega, L)_{\alpha \rightarrow \beta}$ for resonance positions $\epsilon_r/eV=-0.3,0,0.3$ 
	and $\gamma/eV=0.01$ are shown. 
    Shifting the resonance or the potentials will 
	change the positions of the steps and the impact of the contributions.}
\end{center}
\end{figure}
Here the $J_l$ are the Bessel functions of the first kind. The dimensionless parameters
$\alpha_L = \frac{a + 1}{2}\alpha$  and $\alpha_R = \frac{a - 1}{2}\alpha$ define the 
strength of the ac-drive in the contacts via $\alpha=\frac{eV_{ac}}{\hbar \omega}$ and 
the asymmetry parameter $a \in [-1,1]$. $V_{ac}$ denotes the amplitude of the ac-bias 
coupling to the DB system and $\omega$ the corresponding driving frequency.
In order to write down the current one has to integrate the expression for the current 
operator and replace the statistical averages of the creation and annihilation operators 
by their equilibrium values. For the Fermi function in lead $m$ the abbreviation 
$f^{\mathrm{e}}_{m}(\epsilon) = \left[\exp\left(\beta_m (\epsilon - \mu_{m})\right)+1\right]^{-1}$ with 
$\beta_m=1/(k_B \mathcal{T}_m)$  is used. Unoccupied states, in other words occupied hole-like states,
 are denoted by $f^{\mathrm{h}}_{m}(\epsilon) =1-f^{\mathrm{e}}_{m}(\epsilon)$.\\
We treat our setup as a Fabry-P\'erot like DB system, for which the transmission probability 
$T(\epsilon, \epsilon_r)=t^{\dagger}(\epsilon, \epsilon_r) t(\epsilon, \epsilon_r)$ 
is well known~\cite{Datta}. 
Incoming and outgoing scattering states are related by the energy-dependent scattering matrix 
(s-matrix) via $\hat{b}_{n}(\epsilon)=\sum_{m} s_{mn}\hat{a}_m(\epsilon)$. 
The s-matrix is of the form 
\begin{equation}
s(\epsilon, \epsilon_r)=\left(
	\begin{array}{cc}
	r(\epsilon, \epsilon_r) & t'(\epsilon, \epsilon_r) \\
	t(\epsilon, \epsilon_r) & r'(\epsilon, \epsilon_r)
	\end{array} 
	\right)	\,.
\end{equation}
 For a resonant level we can use the Breit-Wigner expression to define the matrix elements 
and thus the transmission through the scattering region via
\begin{equation}
s_{m n}(\epsilon, \epsilon_r) = 
\delta_{m n} - i\frac{\sqrt{\gamma_{m} \gamma_{n}}}{\epsilon -\epsilon_r  + i\frac{\gamma}{2}} \; \text{,} \label{BreitWigner}
\end{equation}
where $\gamma=\sum_{m} \gamma_{m}$ is the half width of the resonance and
$\epsilon_r$ is the resonance energy of the level. 
In general the barrier strength $\gamma_n$ could also depend on energy which  
we neglect here for simplicity. Furthermore we assume 
symmetric barriers $\gamma_L = \gamma_R = \gamma/2$ and call the setup symmetric if  
the resonance is at the Fermi energy ($\epsilon_r=0)$ and
 $-\mu_L = \mu_R = eV/2$.

 If the s-matrix does not dependent on energy, quantum-noise generated
 by the current partitioning at the scattering region can be traced
 down to fluctuations in the electronic occupations of the contact
 with the emission of carriers from left and right
 leads~\cite{MartinLandauer92}.  These fluctuations are the sum of
 variances of the possible current pulses of incident (or empty) wave
 packets at left and right contacts times their weight factors.  An
 incident wave packet can either be transmitted, with probability $T$,
 or reflected with probability $1-T$. It has been shown that in this
 limit completely closed ($T=0$) or open ($T=1$) channels can not
 produce any noise, since either no charge is transferred or their is
 no partitioning at the scatterer.  For intermediate values of $T$ the
 quantum noise in this regime consists of four linear contributions.
 Two contribution with initial and final states related to the same
 terminal with onsets at $\Omega=0$ and two contributions with initial
 and final states at opposite terminals and onsets at $\hbar \Omega
 =\pm eV$. This limit is approached in the spectrum of
 Fig.~\ref{fig-1} for $\gamma\gtrsim eV$.  Thus, at zero temperature
 the asymmetric noise spectrum is nonzero if $\hbar \Omega > -eV$ and
 exhibits kinks at frequencies $\hbar \Omega=0, eV$.  When performing
 the zero-frequency limit some contributions will be absent due to
 Pauli principle. This is e.g the case for current pulses incident
 from right and left lead where one is transmitted and the other one
 reflected, the whole process being proportional to $T(1-T)
 f^{\mathrm{e}}_{\alpha}(\epsilon)f^{\mathrm{e}}_{\beta}(\epsilon)$,
 because then
 $f^{\mathrm{e}}_{\alpha}(\epsilon)=f^{\mathrm{e}}_{\beta}(\epsilon)$.
 However, at finite frequency and with additional ac-driving it is in
 general not possible to express the noise in terms of transmission or
 reflection probabilities but one has to interpret the different
 products of s-matrices involved in the four contributions to the
 noise.  The weight of these contributions is given by the
 Besselfunctions $J_n\left(\alpha\right)$ that describe a
 photon emission or absorption processes of order $n$ at driving
 strength $\alpha$. The noise spectral density is defined as
\begin{widetext}
\begin{align}
 &S_{\alpha \beta}(\Omega,\omega) = \left(\frac{e^2}{2 \pi \hbar}\right) \int d\epsilon \underset{\gamma \delta, l k m}\sum
J_l\left(\frac{eV_{\gamma} }{\hbar \omega} \right) J_k\left(\frac{eV_{\delta} }{\hbar \omega} \right) J_{m+k-l}\left(\frac{eV_{\delta} }{\hbar \omega} \right) J_m\left(\frac{eV_{\gamma} }{\hbar \omega} \right) \nonumber\\
&  Tr\left[{\bf A}_{\gamma \delta}(\alpha, \epsilon, \epsilon + \hbar \Omega) 
{\bf A}_{\delta \gamma}(\beta, \epsilon + \hbar \Omega + (m-l) \hbar \omega, \epsilon+ (m-l) \hbar \omega, ) \right]
f^{\mathrm{e}}_{{\gamma}}(\epsilon-l \hbar \omega) f^{\mathrm{h}}_{{\delta}}(\epsilon +\hbar \Omega -k \hbar \omega) \label{patnoise}
\end{align}
\end{widetext}
With the so-called current matrix 
${\bf A}_{\gamma \delta}(\alpha, \epsilon, \epsilon')=\delta_{\alpha \gamma} \delta_{\beta \delta} - s^*_{\alpha \gamma}(\epsilon) s_{\beta \delta}(\epsilon')$ which connects incoming and outgoing states via 
the s-matrices at different energies.
\begin{figure}[t]
  \begin{center}
    \includegraphics[angle=270,width=0.94\columnwidth,clip]{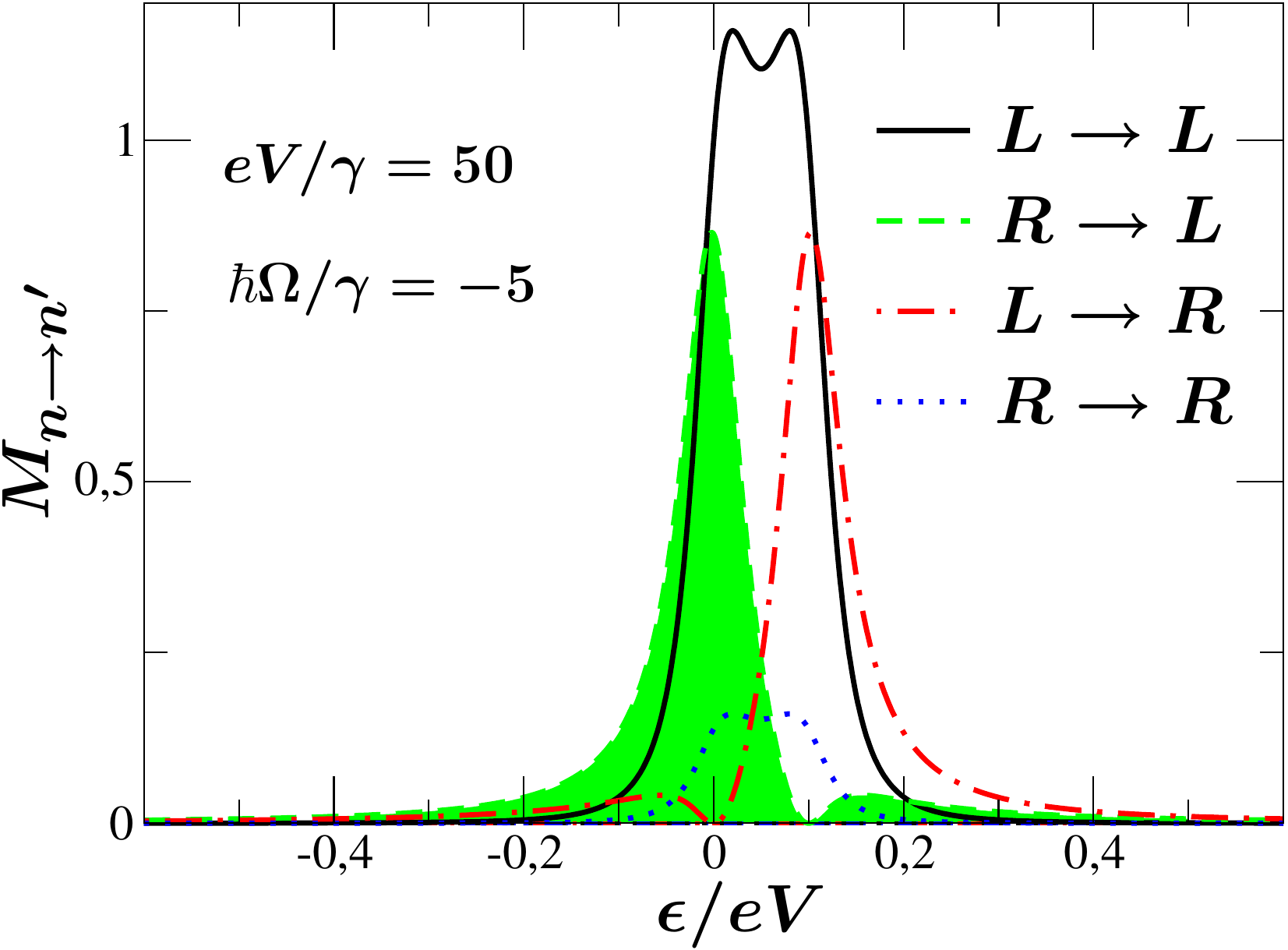}\\
    \includegraphics[angle=270,width=0.94\columnwidth,clip]{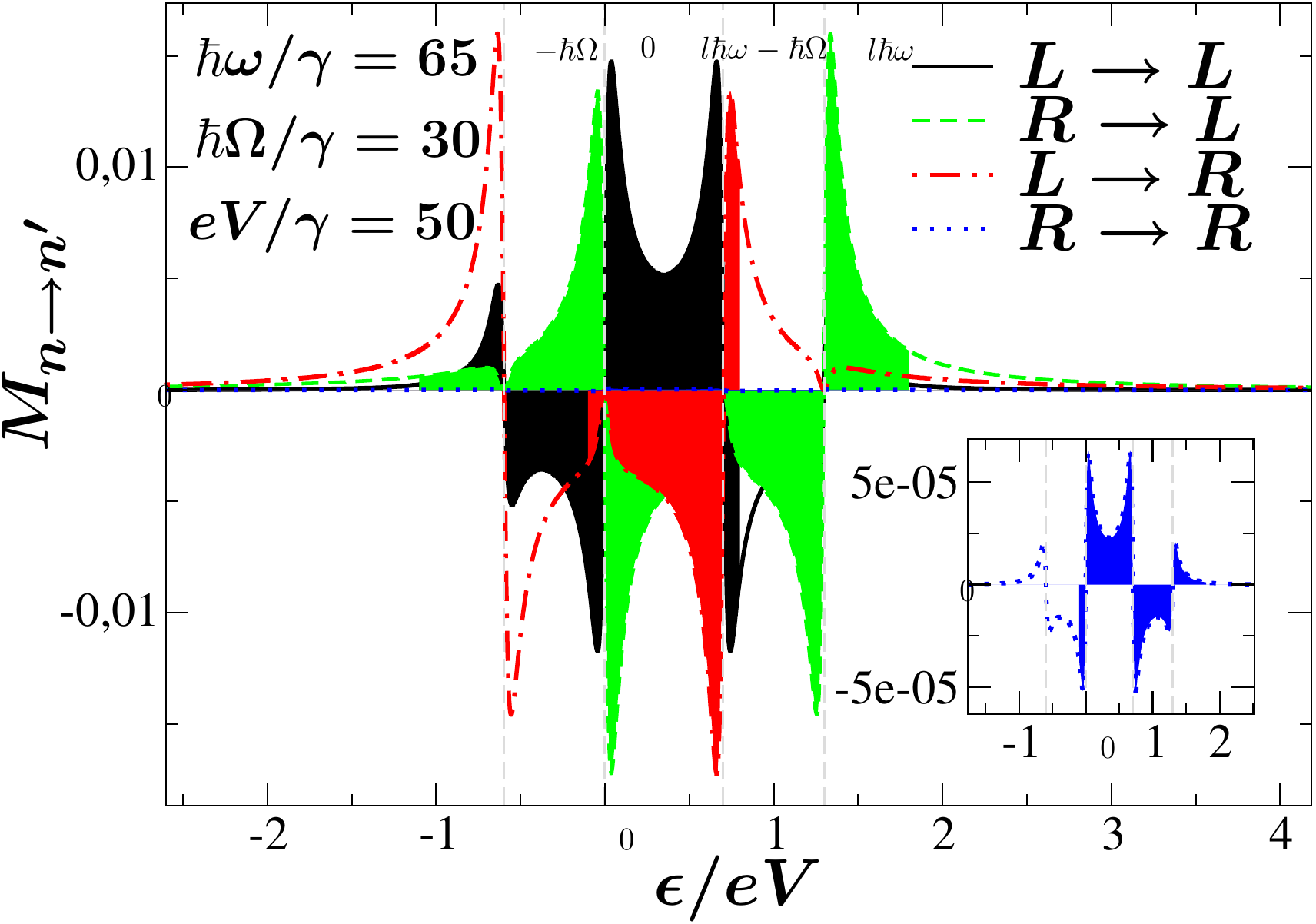}
    \caption{
      \label{fig-2} 
      Examples of the integrands real parts for finite $eV$ without
      (top) and with (bottom) ac-driving.  We show the integrands
      $M_{n \rightarrow n'}(\Omega, \omega)$ with $(n,n')=$ (L,L),
      (L,R), (R,L) and (R,R) as a function of $\epsilon$ for a
      centered resonance ($\epsilon_r=0$).  The lines show the
      integrand while the filling denotes the integration interval.
      Top: No ac-driving, so at negative frequency only $R \rightarrow
      L$ contributions are present, as indicated by the shaded region.
      The $L \rightarrow R$ term is similar to the other
      cross-terminal contribution, both have a single peak, located at
      $\epsilon=\epsilon_r$ or at $\epsilon=\epsilon_r-\hbar
      \Omega$. If $\Omega \rightarrow 0$ these terms interfere
      destructively leading to a double peak shape with maximal values
      of $1/4$ around the local minima at $\epsilon=0,-\hbar \Omega$.
      The auto-terminal contributions have a double-peak structure at
      the same energies. For $\Omega \rightarrow 0$ the $L\rightarrow
      L$ term can be $>1$, but it is not probed for a centered
      resonance. Its significant contributions are at $\hbar \Omega
      \ge eV/2$ and not shown in the example.  Bottom: Integrands of
      the ac-driven setup with $\hbar \Omega/\gamma=30$, $\hbar
      \omega/\gamma=65$, and $l=1$; other parameters as above.  Due to
      the complex s-matrices negative values are possible.}
\end{center}
\end{figure}
If one of the frequencies involved is zero at least some correlators
can be written in terms of $T(\epsilon)$ and $R(\epsilon)$. But in
general this is not the case due to the special role of the complex
reflection amplitudes.  In equilibrium ($eV=0,\alpha=0$) these
amplitudes lead to finite noise even if no transmission through the
system is possible~\cite{Buttiker92}. We will emphasize their special
role concerning the noise spectral function if finite bias voltages
are applied.  Therefore we separate the dc-noise spectrum into a sum of
states which are scattered from terminal $\alpha$ to terminal $\beta$:
\begin{equation}
	S_{LL}(\Omega,\omega):=\underset{\alpha, \beta =L,R}\sum   C^{\mathrm{}}_{\alpha \rightarrow \beta}(\Omega,\omega) \label{SLLsum}
\end{equation}

The four correlators contributing to auto-correlation noise without time-dependent voltages ($\omega=0$) are then determined by
\begin{subequations}
	\label{SLL0auto}
\begin{align}
	C^{\mathrm{}}_{L \rightarrow L}(\Omega)=& 
	\frac{ e^2}{2 \pi \hbar} \Theta(\hbar \Omega) \underset{\mu_L-\hbar \Omega}{\overset{\mu_L}\int}  d\epsilon \,
	 \left|r^*(\epsilon)r(\epsilon+\hbar\Omega) - 1 \right|^2  \label{correlators:LtoL}\\
%%%%%%%%%%%%%
	C^{\mathrm{}}_{R \rightarrow R}(\Omega)=&
	\frac{e^2}{2 \pi \hbar}  \Theta(\hbar \Omega)\underset{\mu_R-\hbar \Omega}{\overset{\mu_R}\int}  d\epsilon \, 
 	T(\epsilon)T(\epsilon+\hbar\Omega)\label{correlators:RtoR}\\
%%%%%%%%%%%%%
	C^{\mathrm{}}_{L \rightarrow R}(\Omega)=&
	 \frac{e^2}{2 \pi \hbar} \Theta(\hbar \Omega-eV)\underset{\mu_R-\hbar \Omega}{\overset{\mu_L}\int}  d\epsilon \, 
 	R(\epsilon)T(\epsilon+\hbar\Omega)\label{correlatorsL:toR}\\
%%%%%%%%%%%%%
	C^{\mathrm{}}_{R \rightarrow L}(\Omega)=& 
	\frac{e^2}{2 \pi \hbar} \Theta(\hbar \Omega+eV) \underset{\mu_L-\hbar \Omega}{\overset{\mu_R}\int}  d\epsilon \,
	T(\epsilon)R(\epsilon+\hbar\Omega)\,. \label{correlators:RtoL} 
\end{align}
\end{subequations}
\begin{figure}[t]
\begin{center}
\includegraphics[width=0.94\columnwidth]{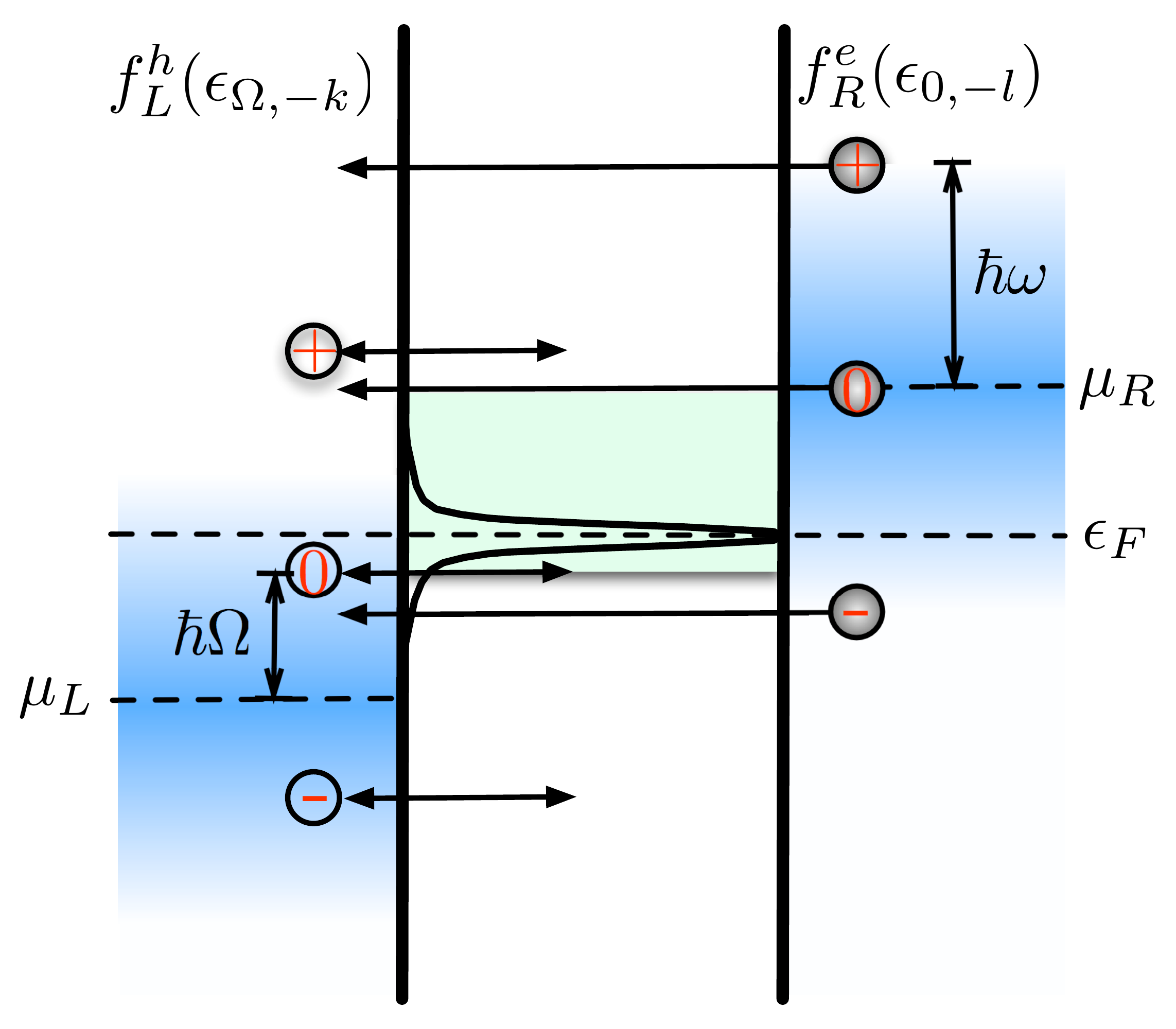}     
\caption{
		\label{fig-3} 
		Elementary events of $C^{\mathrm{}}_{\alpha \rightarrow \beta}(\Omega, \omega)$. We have chosen 
	    $(\alpha,\beta) =$ (R,L) and $\epsilon_r=0$ as an example. 
	    The frequency is fixed close to the step at $\hbar \Omega = -eV/2$
	 	in the noise spectrum. A higher frequency would shift the lower bound of the integration window (colored region around the resonance) 
		and the hole-like final states downward towards $\mu_L$. The filling of the left 
		and right reservoirs refer to the applied dc- and ac-bias voltages. 
		The available free and occupied states are defined by the Fermi functions for 
		electrons $f^{\text{e}}_{L/R}(\epsilon_{\Omega},k)=f_{L/R}(\epsilon+\hbar \Omega+k\hbar \omega )$ and for holes
		$f_{L/R}^{\text{h}}(\epsilon_{\Omega},k)=1-f_{L/R}(\epsilon+\hbar \Omega+k\hbar \omega )$. 
		Arrows indicate the possible mechanism which are suggested by the products of s-matrices that appear in the integrands.
		We show the contributions up to first order in the driving $n\hbar \omega$, so $n=0,\pm1$.
		The colored region in-between the barriers denotes the integration interval when n=0.
		}  
\end{center}
\end{figure}
Here the correlator of Eqn.~(\ref{correlators:LtoL}) can not
explicitly be written as a product of probabilities. Rather we find a
term with states scattered from and back to lead $L$ describing the
two-particle quantum interference of coherently scattered
quasiparticles with the occupied states in the lead where
current-fluctuations are measured.  The quasiparticles in the lead can
interfere with either a reflected quasi-electron that absorbs a quanta
$\hbar \Omega$ or with a quasi-hole propagating along the inverse path
and emitting a photon with energy $\hbar \Omega$. In terms of
probabilities $C^{\mathrm{}}_{L \rightarrow L}(\Omega)$ acquires a
finite scattering-phase $\Phi(\epsilon,\Omega) =
\text{Arg}\left[r^*(\epsilon)r(\epsilon+\Omega)\right]$ via its
integrand that can be written as $(1 +
R(\epsilon)R(\epsilon+\hbar\Omega) - 2 [R(\epsilon)R(\epsilon+\hbar
\Omega)]^{1/2}
\cos(\Phi(\epsilon,\Omega)))f^{\mathrm{e}}_{L}(\epsilon)f^{\mathrm{h}}_{L}(\epsilon+\hbar
\Omega)$.  Moreover it is this contribution that can produce noise
even for vanishing transmission, in analogy to the equilibrium
problem. For our choice of chemical potentials the only non-vanishing
correlator at zero-frequency is given by Eq.~(\ref{correlators:RtoL}).

Without ac-bias voltages but at finite frequency the auto-correlations
are real and the cross-correlations at opposite terminals are the
hermitian conjugate of each other, so they obey the symmetries:
\begin{align}
S^{\dagger}_{LL}(\Omega) &= S_{LL}(\Omega)\\
S^{\dagger}_{LR}(\Omega) &= S_{RL}(\Omega)
\end{align}
In addition, if $\Omega=0$, the well-known symmetry $S_{\alpha
  \alpha}(\Omega=0)=-S_{\alpha \beta}(\Omega=0)$ is recovered, so the
sum of all current-correlations vanishes
$\sum\limits_{\alpha,\beta=L,R} S_{\alpha\beta}(\Omega=0)=0$, see also
Refs.~\cite{BlanterButtiker00,Rothstein08}.
In order to develop an intuitive interpretation for products of two
s-matrices $s^*_{\alpha \beta}(\epsilon)s_{\alpha'
  \beta'}(\epsilon+\hbar \Omega)$ we express them in terms of
probabilities. If both s-matrices have the same indices
$\alpha'=\alpha$ and $\beta'=\beta$ we introduce the transmission and
reflection functions
\begin{align}
T(\epsilon, \epsilon+\hbar \Omega)&= t^*(\epsilon) t(\epsilon + \hbar \Omega)\\
R(\epsilon, \epsilon+\hbar \Omega)&=	r^*(\epsilon) r(\epsilon + \hbar \Omega)\,.
\end{align}
In terms of the usual probabilities $T(\epsilon)$ and $R(\epsilon)$ we find
\begin{align}
&T(\epsilon, \epsilon+\hbar \Omega)=\nonumber\\
&T(\epsilon) \left(\frac{\epsilon+\hbar \Omega \,T(\epsilon+\hbar \Omega)}{\epsilon + \hbar \Omega}	+ \frac{i \hbar \Omega \,T(\epsilon+\hbar \Omega)}{\gamma} \right)  \\
&R(\epsilon, \epsilon+\hbar \Omega)=\nonumber\\
&R(\epsilon)\left(\frac{\epsilon+\hbar \Omega \,T(\epsilon+\hbar \Omega)}{\epsilon}+	\frac{i \hbar \Omega(\epsilon+\hbar \Omega) T(\epsilon +\hbar \Omega)}{\epsilon \gamma}\right) \,.	\label{transmission}			
\end{align}
For $\Omega \rightarrow 0$ these expressions reproduce the probabilities $T(\epsilon)$ and $R(\epsilon)$.  
At finite $\Omega$ they illustrate nicely how an imaginary part and at the same time an additional contribution to
the real part are acquired, both proportional to $\hbar \Omega \, T(\epsilon + \hbar \Omega)$. At the same time contributions 
proportional to the probability $T(\epsilon)$ are modified by a factor $\epsilon/(\epsilon+\hbar \Omega)$.
Depending on the value of $\Omega$ this can lead to a reduced or enhanced transmission function for those processes.
The imaginary part can be seen as a finite scattering time in the FP setup where the 
corresponding timescale is given by the inverse resonance width $1/\gamma$.
If we allow arbitrary pairings of s-matrices at energies separated by the frequencies $\Omega,\omega$, as they appear in
in Eq.~(\ref{patnoise}) for the noise spectral function with finite ac-bias voltage,
we find the transmission functions
%
%\begin{widetext} 
%
\begin{subequations}
	\label{transmissionFunctions}
	\begin{align}
 &T(\Omega_m,\omega_n):=s^*_{LR}(\Omega_m)s_{LR}(\omega_n)\nonumber \\
	& = T(\Omega_m) T(\omega_n)\left( 1 + \frac{\epsilon_m \epsilon_n }{\gamma^2} + i \frac{\omega_n - \Omega_m}{\gamma} \right) \\
& R(\Omega_m,\omega_n):=s^*_{LL}(\Omega_m)s_{LL}(\omega_n)\nonumber\\
	&= R(\epsilon_m) R(\epsilon_n)\left(1+ \frac{\gamma^2}{\epsilon_m \epsilon_n} + i \frac{\gamma (\omega_n-\Omega_m)}{\epsilon_n \epsilon_m}\right)  \\
&	M(\Omega_m,\omega_n):= s^*_{LL}(\Omega_m)s_{LR}(\omega_n)\nonumber\\
	& = R(\Omega_m) T(\omega_n) \left(\frac{\omega_n-\Omega_m}{\epsilon_n}  + i  \frac{\epsilon^2_n + \gamma^2}{\gamma \epsilon_n}\,. \right)
\end{align}
Above we used the shorthands $\omega_m  (\Omega_m)=m \hbar \omega (\Omega)$ and $\epsilon_{n(m)}=\epsilon+ \omega_n (\Omega_m)$, with integer $m,n$.
\end{subequations}

Since we only regard symmetric coupling to the leads ($\gamma_L=\gamma_R$) the s-matrices are 
invariant when exchanging reservoir indices $L$ and $R$. 
Then the noise is symmetric under exchange of
the indices $L,R$ if the dc-bias is reversed, too. Therefore we only deal with the 
auto-correlation and cross-correlation noise $S_{LL}(\Omega, \omega)$ and $S_{LR}(\Omega, \omega)$. Consequently we also give the formulas in terms of $t(\epsilon)=s_{LR}(\epsilon)=s_{RL}(\epsilon)$ as well as $r(\epsilon)=s_{LL}(\epsilon)=s_{RR}(\epsilon)$.  
\section{Current-current auto correlations}
The description in terms of initial and final states defined by the Fermi function products
is supported by expressing the noise spectrum with the help of  Fermi's golden
rule \cite{GavishImry00,GavishImry01}: 
\begin{equation}
S_{\alpha \alpha}(\Omega)= 2\pi \underset{\mathrm{i,f}}\sum P_{\mathrm{i}} \left| \left<\mathrm{i} \left|\Delta\hat{I}_{\alpha} \right| \mathrm{f} \right> \right|^2 \delta\left(\epsilon_{\mathrm{i}} - \epsilon_{\mathrm{f}}  - \hbar \Omega \right)  \, ,
\end{equation}
where $P_{i}$ is the probability that the initial state is filled, here described by the grand-canonical ensemble. 
The system absorbs photons $\hbar \Omega$ from an electric field and tunnels from the 
initial state $\left|{\mathrm{i}} \right>=\left|i,n \right>$ with $n$ photons to the 
final state $\left|{\mathrm{f}} \right>=\left|f, n+1 \right>$ containing $n+1$ photons. 
In the same way the substitution $\Omega \rightarrow -\Omega$ describes emission of photons with final states containing $ n-1$ photons.
Then the sum of emission and absorption processes can be used to relate the noise spectrum to the ac-conductivity.
For a Breit-Wigner lineshape, Eq.~(\ref{BreitWigner}), the noise spectral density can be calculated analytically at $k_B\mathcal{T}=0$.
In the dc-limit integration of Eqs.(\ref{SLL0auto}) yields 
\begin{subequations}
	\label{eqn:SLL0autoResults}
	\begin{align} % CLtoL auto dc
		C_{L \rightarrow L}(\Omega,V)& =\Theta(\Omega)f(\Omega)(1+(\Omega/\gamma)^2)F(\mu_L-\epsilon_r,\Omega)\\
		C_{R \rightarrow R}(\Omega,V)& =\Theta(\Omega)f(\Omega)F(\mu_R-\epsilon_r,\Omega)\\
		C_{L \rightarrow R}(\Omega,V)&=\Theta(\Omega-eV)f(\Omega)\left.G(\epsilon-\epsilon_r,\Omega)\right|^{\mu_L}_{\mu_R-\Omega}\\
		C_{R \rightarrow L}(\Omega,V)&=\Theta(\Omega+eV)f(\Omega)\left.G(\epsilon-\epsilon_r,\Omega)\right|^{\mu_R}_{\mu_L-\Omega}\,,
	\end{align}
\end{subequations}
where we used the definitions provided in the appendix, Eqs.~(\ref{app:def_f})-(\ref{app:def_AB}).
The result for $C^{\mathrm{}}_{R \rightarrow L}(\Omega,V)$ is identical to $C^{\mathrm{}}_{R \rightarrow L}(\Omega,V)$ when we interchange the reservoir indices $L,R$ and thus replace $eV$ by $-eV$ in the pre-factor. When the setup is symmetric 
the result for the cross-terminal contributions is defined by the replacement 
$\left.G(\epsilon-\epsilon_r,\Omega)\right|^{\mu_L}_{\mu_R-\Omega}\rightarrow H(-V/2, \Omega)$ and $\left.G(\epsilon-\epsilon_r,\Omega)\right|^{\mu_R}_{\mu_L-\Omega}\rightarrow H(V/2, \Omega)$.
Obviously, the unique fingerprint of the terminal $L$, where the fluctuations are probed, is given by the  additional frequency-dependence in the pre-factor.
Moreover, at $\Omega=0$ the noise power is defined by $S_{LL}(0,V)= C^{\mathrm{}}_{R \rightarrow L}(0,V)$ where
\begin{align}
	  C^{\mathrm{}}_{R \rightarrow L}(0,V) & 
          =\frac{e^2 \gamma }{4\hbar}
          \left[\mathrm{atan}\left(\frac{\mu_R-\epsilon_r}{\gamma
              }\right)- 
            \mathrm{atan}\left(\frac{\mu_L-\epsilon_r}{\gamma }\right)\right.\nonumber \\ 
&\left.+\gamma \left(\frac{\mu_R-\epsilon_r} {(\mu_R-\epsilon_r)^2+\gamma ^2}-\frac{\mu_L -\epsilon_r}{(\mu_L-\epsilon_r)^2+\gamma ^2}\right) \right]\,.
\end{align}
Thus, at $-\mu_L=\mu_R=eV/2$ and with $\gamma \ll eV$ we have $C^{\mathrm{}}_{R \rightarrow L}(0,V)=e^2\gamma \pi/2\hbar $. This results in the well known  sub-Poissonian  Fano factor $F\equiv S_{LL}(\Omega=0,\omega)/eI=1/2$. In the opposite limit, when $\hbar \Omega \gg eV$, the correlators approach the values
\begin{subequations}
	\begin{align}
		C^{\mathrm{}}_{L \rightarrow L}(\Omega\rightarrow \infty,V)&=\frac{e^2 \pi \gamma}{2\hbar }\\
		C^{\mathrm{}}_{R \rightarrow R}(\Omega\rightarrow \infty,V)&=0\\
		C^{\mathrm{}}_{L \rightarrow R}(\Omega\rightarrow \infty,V)&=\frac{e^2\gamma}{2 \hbar}\left[\pi -2 \mathrm{atan}\left[\frac{eV}{2\gamma}\right] \right]\\
	    C^{\mathrm{}}_{R \rightarrow L}(\Omega\rightarrow \infty,V)&=\frac{e^2\gamma}{2 \hbar}\left[\pi+2 \mathrm{atan}\left[\frac{eV}{2\gamma}\right] \right]  \,,
	\end{align}
\end{subequations}
\begin{figure}[t]
\begin{center}
\includegraphics[width=0.99\columnwidth]{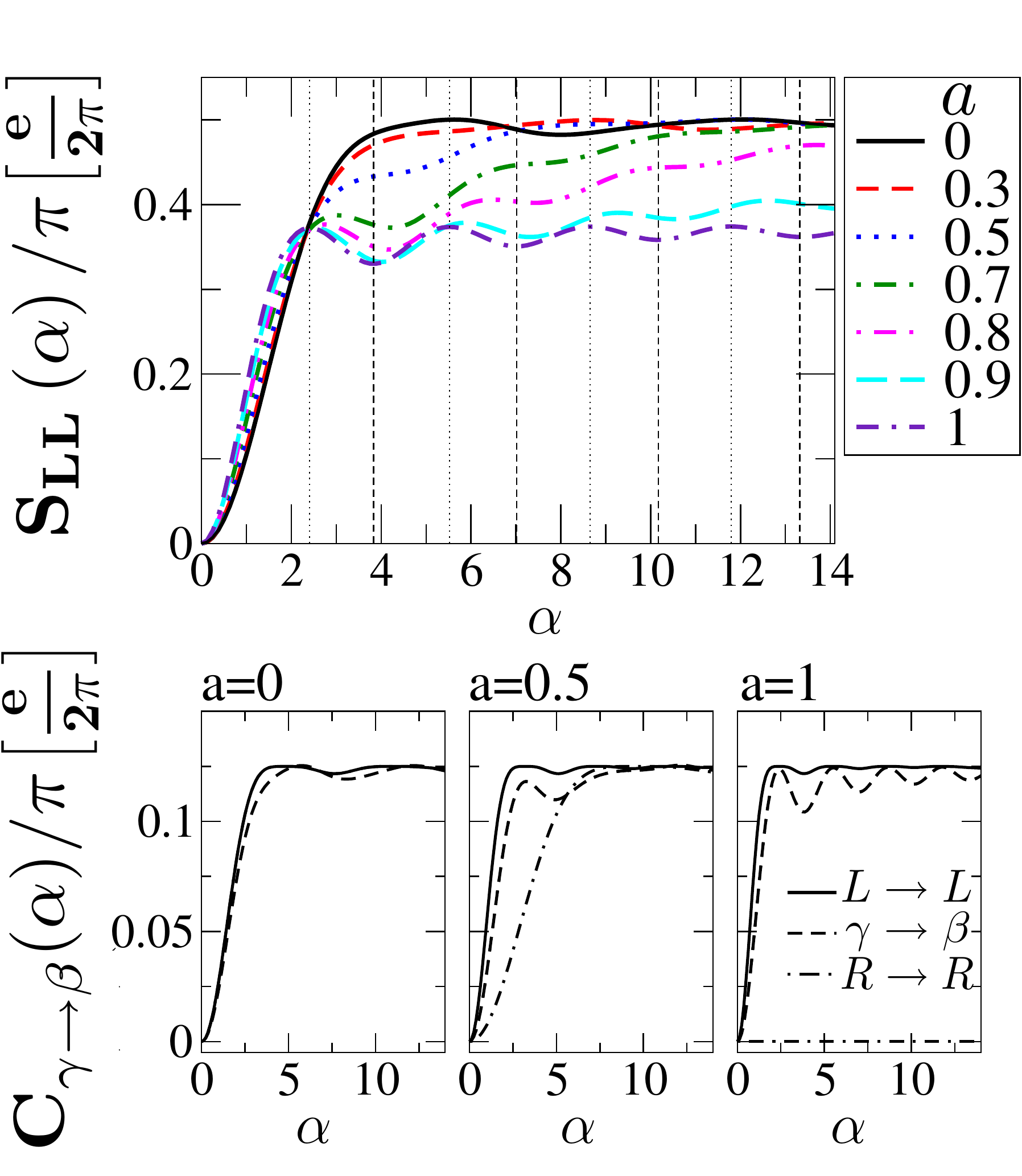}
\caption{
\label{fig-4} Zero dc-bias auto-correlations but finite ac-bias. 
				Top: Noise vs driving for different values of the asymmetry parameter $a$. Other parameters are 
				$\hbar \omega/\gamma=50$, $\alpha=1.86$, $eV,\Omega=0$.  
				Bottom: Some examples ($a=0,0.5,1$) of the four contributions to the noise vs $\alpha$.         
				 For $a\rightarrow 1$ ($V_{ac,R}=0$) the $R \rightarrow R$ term vanishes (bottom).
				Then $S^{\mathrm{}}_{LL}(\hbar \Omega=0,eV=0)$
				oscillates around smaller values in comparison to the $a=0$ configuration. 
				}
\end{center}
\end{figure}
\begin{figure}[t]
\begin{center}
\includegraphics[width=0.99\columnwidth]{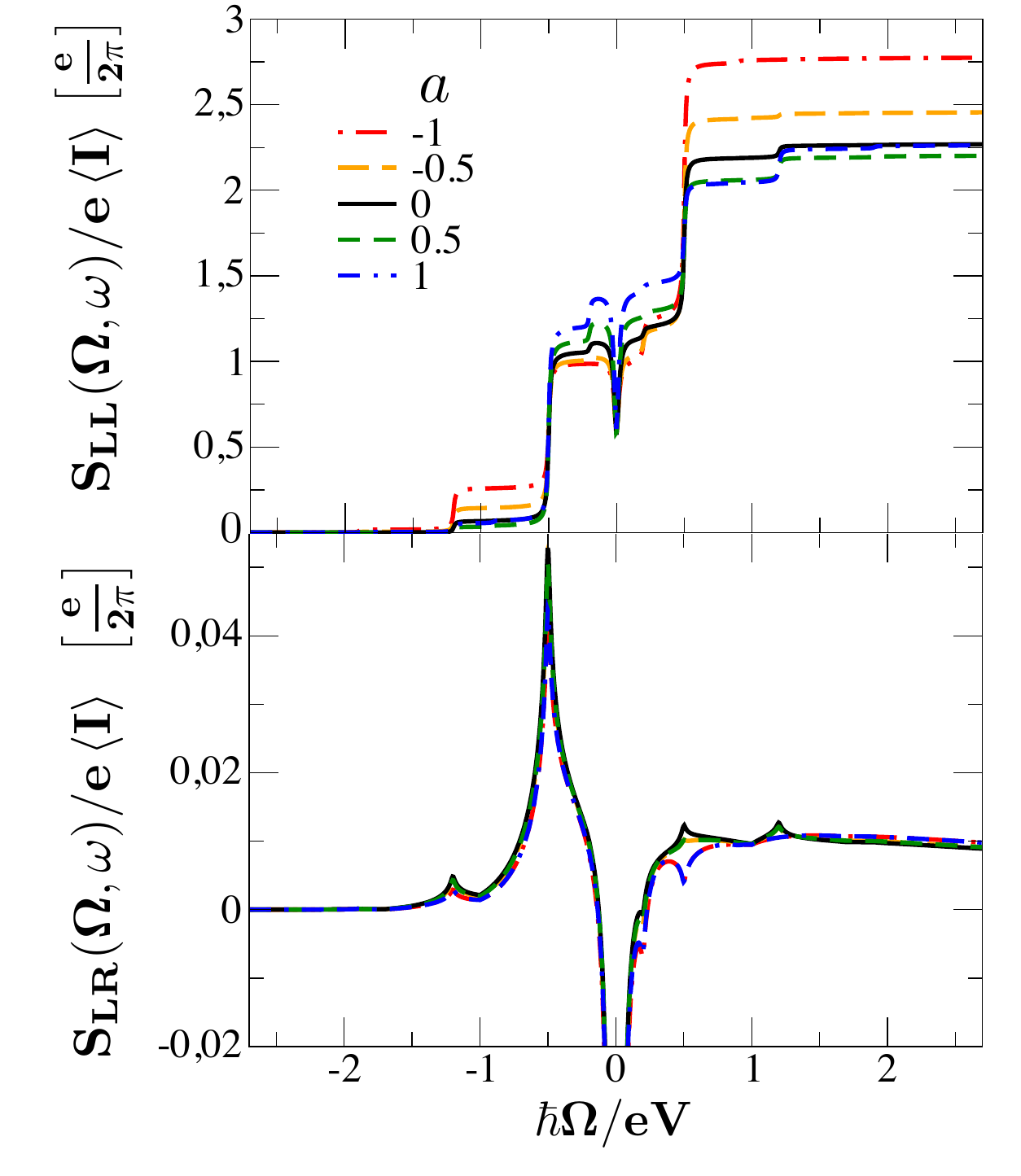}
\caption{
\label{fig-5} Auto-correlation (top) and cross-correlation noise spectral density (bottom) with 
				ac-driving ($\alpha=1.86$, $\hbar \omega/\gamma=35$, $a=1$). The coincidence of steps in the auto-terminal 
				and peaks in the cross-terminal noise spectra is only partial, since the integrand
				 in dominating contributions probes peaks at different energies. 
				The (local) minimum at $\hbar \Omega=0$ of the (auto) 
				cross-terminal noise translates for the chosen parameters into a Fano factor $F=(+)- 0.5$.  
}
\end{center}
\end{figure}
in agreement with Fig.~\ref{fig-1}. For large bias voltages $C^{\mathrm{}}_{L \rightarrow R}(\Omega\rightarrow \infty,V)=0$ whereas $C^{\mathrm{}}_{R \rightarrow L}(\Omega\rightarrow \infty,V)$ and  $C^{\mathrm{}}_{L \rightarrow L}(\Omega\rightarrow \infty,V)$ both contribute to the frequency-dependent Fano factor with unity. Thus, for large frequencies the Fano factor approaches $F=2$. Due to the lengthy expressions that occur when finite ac-bias is applied,
we provide the analytical results in the appendix, Eqs.(\ref{app:SLLac_results}). 
Then Fano factors $F>2$ are possible since the average dc-current can be suppressed by the ac-bias voltage.
Besides the onsets of the correlators and their interpretation in terms of absorption ($\Omega > 0$) and 
emission ($\Omega < 0$) of photons by the scattered quasiparticles there is a second important ingredient
that determines the current fluctuations. Namely, if the energy is provided there has to exist a scattering channel
so a quasiparticle can contribute to the current and current-noise. This is determined by the integrand, the 
distance of the resonant level to the chemical potentials of the reservoirs and the resonance width.
The interplay of these features will be discussed in the following intensively. 
\subsection{Effect of finite frequency}
In the noise spectrum of Fig.~\ref{fig-1}
the first step of $S^{\mathrm{}}_{LL}(\Omega)$ is determined by states contributing 
to $C^{\mathrm{}}_{R\rightarrow L}(\Omega)$. For a centered resonance the distance of the 
resonance to the chemical potential of the left reservoir is $-eV/2$,
so the step is at the corresponding frequency. If we increase the distance to the reservoir 
of the final state the step is shifted to smaller frequencies so the plateau gets wider. 
This behavior can be understood by an argument provided 
by the structure of the involved product of s-matrices (Fig.~\ref{fig-2}). 
This product exhibits a single peak at $\epsilon_r- \hbar \Omega$ that is only probed 
by the noise if it is inside the energy window $\mu_L- \hbar \Omega \ldots \mu_R  $ and a small 
shoulder for energies larger than  $\hbar \Omega$. 
It is clear from above arguments that the step-width is $2 \gamma$. Since for 
frequencies $-eV/2<\hbar \Omega < eV/2$ no further scattering paths exist, the noise 
stays constant in this regime apart from the dip around $\hbar \Omega=0$. 
This sub-Poissonian Fano factor can be understood as the effect of electron anti-bunching. 
Since incoming wave packets hit the scattering region with a rate $1/eV$ and have a temporal 
extension proportional to $1/\gamma$, a frequency $\hbar \Omega \sim \gamma$ can not probe 
the correlation between them. This picture is supported by the fact that at frequencies 
of the order of the resonance width the two correlated  events, which are suggested by the 
Fermi functions, are both in resonance. Namely an electron-like state 
$f_R^{\mathrm{e}}(\epsilon)$ transmitted $T(\epsilon)$ at energy $\epsilon$ from  right to left, 
and a hole state $f_L^{\mathrm{h}}(\epsilon + \hbar \Omega)$, reflected at left terminal at 
energy $\epsilon + \hbar \Omega$ with probability $R(\epsilon + \hbar \Omega)$.
So the integrand in $C^{\mathrm{}}_{R\rightarrow L}$ is suppressed 
by a factor of $2$ (we have a second resonant path) in terms of the 
interference-like dip around $\epsilon=0$. In the mentioned regime the 
transmission can still be aligned with the resonance energy leading to the same charge 
transfer as for higher frequencies, whereas the reflected path is strongly suppressed 
since $R(\epsilon) \rightarrow 0$ as $\epsilon \rightarrow \epsilon_r$, so the ratio $S/eI$ 
should be suppressed. A similar discussion of the other contributions is straight forward. 
The main aspects are: 
The dominating contributions are those where the final state is related to the measurement terminal. 
This is also the terminal where charge is effectively transferred to. 
If the energy transferred via PAT events matches the distance of the resonance 
energy $\epsilon_r$ from the chemical potential $\mu_L=-eV/2$, 
then the interference-like term  $C^{\mathrm{}}_{L\rightarrow L}$ 
leads to the second step, located at $\hbar \Omega =eV/2$ in the noise spectrum. 
Assuming a centered resonance, the integrand for this term exhibits peaks at
$\epsilon=0, -\hbar \Omega$. Those peaks unite to a single one when $\hbar \Omega \le 2 \gamma$ 
(see the black curve in Fig.~\ref{fig-2} a) 
and show destructive interference corresponding to the 
afore mentioned anti-bunching of the quasiparticles. 
If $\mu_L=\mu_R=0$ this behavior is the origin of a small overshoot in 
the auto-correlation spectrum at frequency $\Omega \le \gamma$ before the spectrum
 saturates (not shown in the plots). 

If a finite dc-bias is applied, then the peak around $\epsilon=0$ is outside the 
integration window. But the center of the second peak comes into play when 
$\hbar \Omega \ge eV/2$, thus we find a step there. The smallest impact on the noise comes from 
$C^{\mathrm{}}_{R\rightarrow R}(\Omega)$ and $C^{\mathrm{}}_{L\rightarrow R}(\Omega)$ since they probe the tail of the resonance only. 
The latter one naturally only has a small impact on current-current correlations 
because a quasiparticle needs to be provided with an energy quantum $\hbar \Omega \ge eV$ 
in order to overcome the potential difference. Therefore the resonance position, 
as long as it is inside the bias-window does not affect the onset of the contribution, 
but modifies the impact on the noise.
\subsection{Influence of harmonic ac-driving}
Here we have to distinguish between differently coupled light fields, 
wether the ac-drive is applied at both or one terminal only.
The ac-bias voltage opens additional scattering paths as illustrated in Fig.~\ref{fig-3} for $C^{\mathrm{}}_{R\rightarrow L}(\Omega, \omega)$ with arbitrary $a$. There, PAT events induced by the ac-bias are considered up to first order.
When $a=1$  all contributions to the auto-terminal noise except 
$C^{\mathrm{}}_{L\rightarrow L}(\Omega, \omega)$ are given by the
set of scattering paths determined by the s-matrices without ac-drive.
In that case the two Besselfunction corresponding to the not driven terminal 
generate a Kronecker delta which assures that the two remaining Besselfunctions 
of the other terminal have the same indices. So the product of all 
Besselfunctions is positive by definition. Furthermore the arguments of the 
s-matrices are independent of the driving frequency because only the 
energies $(m-l)\hbar \omega$ with $m-l=0$ are allowed. 
Thus, for scattering events where  one of the two states is related
to a driven reservoir, either the initial or the final one, 
the ac-driving enters only via the $k\hbar \omega$ or $l \hbar \omega$ terms 
in the argument of the Fermi functions but leave the integrand unchanged. 
Consequently, PAT events that are stimulated by the ac-bias voltage show up
in all correlators even if initial and final states are not related to the driven terminal.
But the number of features that can be identified in the noise spectral function increases
when $|a| \ne 1$.
Now let us take a closer look on Fig.~\ref{fig-4}, where $eV=0$ and $\Omega=0$: 
Starting the analysis with the curve for $a=1$ one can identify the minima and maxima 
of $S_{LL}(\Omega,\omega)$ with the zeros of the $J_n(\alpha)$ when $n=0,1$. The surprising fact that the
oscillations have minima when $J_1$ vanishes is due to the $n=0$ term which has no contribution 
to the noise because it does not probe the peak of the involved integrand. 
But for frequencies larger then $\gamma$
the $n=1$ term does and therefore dominates the charge transport and fluctuations.
 The same reason leads to the maxima when $J_0$ has minima and thereby reduces the weight of the zero-order terms. 
Because of the completeness of the Besselfunctions then the higher order terms, 
which are non-zero if $m=l>0$, have a  stronger weight and the noise is enhanced.
In this example the ac-generated PAT contributions to the noise are related to the left reservoir - the oscillating one - 
so the correlations $C^{\mathrm{}}_{R\rightarrow R}(\Omega,\omega)$ vanish. 
The oscillatory behavior of the Besselfunctions is clearly visible for contributions $C_{L\rightarrow L}(\Omega,\omega)$ 
and for the two cross-contributions to $S_{LL}^{\mathrm{}}(\Omega=0,\omega)$. 
These are identical and exhibit even more pronounced oscillations, with a maximal contribution of 
$C_{\alpha \rightarrow \beta}(\Omega,\omega)/eI = 0.125$.
\begin{figure}[t]
\begin{center}
\includegraphics[width=0.99\columnwidth]{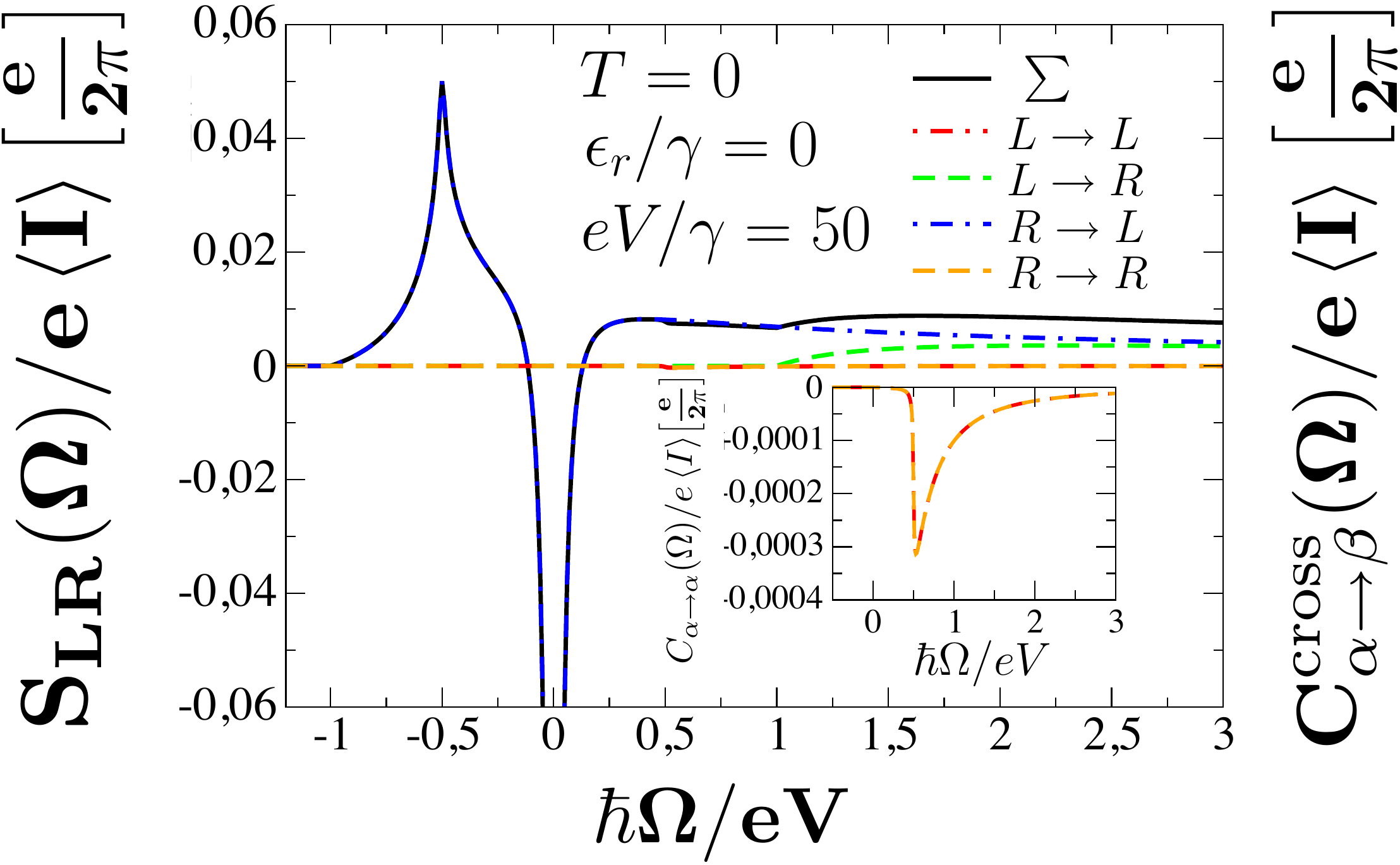}
\caption{
\label{fig-6} Frequency dependence of the zero-temperature current-current cross-correlations 
			for $\epsilon_r = 0$. The minima at $\hbar \Omega \rightarrow 0$ is  given 
			by $\left. S_{LR}=-S_{LL}\right|_{\Omega=0}$. 
		  	Dominating terms inside the 
			integration intervals are originating from the cross-terminal contribution. The inset shows a zoom into auto-terminal terms,
			which are identical ($\epsilon_r=0$) and orders of magnitude smaller than cross-terminal ones due to the sharp resonance considered.
		 	Now a peak-like structure can be observed instead of the step-like behavior observed in the auto-correlation spectrum.}
\end{center}
\end{figure}
That's why the two limiting cases $a=1$ and $a=0$ have maximal values of $0.5$ and $0.375$.
If the asymmetry in the driving is reduced, as done above (meaning we increase the amplitude 
of the driving with opposite sign at the second reservoir), 
the contributions $C^{\mathrm{}}_{R\rightarrow R}(\Omega,\omega)$ are finite. 
As an example we analyse the curve for $a=0.5$. This means in the left reservoir we have 
an effective driving of the order $0.75 \alpha$ while at the right reservoir of the order $0.25 \alpha$. 
Consequently for $C^{\mathrm{}}_{L\rightarrow L}(\Omega,\omega)$ we find maxima where $J_n(0.75 \alpha)$ has minima 
and for $C^{\mathrm{}}_{R\rightarrow R}(\Omega,\omega)$  where $J_n(0.25 \alpha)$ has minima.
Since we are analyzing a situation where $\epsilon_r=0$ (symmetric setup) the two 
cross-contributions $C^{\mathrm{}}_{\alpha\rightarrow \beta}(\Omega,\omega)$ ($\alpha \ne \beta$)
to the auto-correlation noise are identical at $eV=0$, showing minima at intermediate positions between 
the expected minima related to $\alpha_{L}$ and $\alpha_{R}$. Concerning the cross-correlation 
spectrum at $\hbar \Omega=0$ as a function of the driving the results are analogues. 
The curve starts at $S_{LR}(\alpha=0)=0$ and oscillates around negative values between 
$-1/2 \ldots -3/8$ ($a=-1 \ldots 1$). At finite voltage the curves start at $\pm 1/2$ and still 
show the oscillations due to the Besselfunctions. But, e.g. for the auto-correlator 
in our setup, contributions scattered into the left reservoir (the driven one) are again dominant. 
Then the $C_{R\rightarrow L}(\Omega,\omega)$ term  is the one giving the finite value at zero driving, 
consistent with the dc noise spectra. The second dominant contribution at $\hbar \Omega=0$, 
$C_{L\rightarrow L}(\Omega,\omega)$, is switched on by the driving voltage.
When an additional ac-bias voltage is applied, the noise spectral function as plotted in Fig.~\ref{fig-5} acquires additional steps 
due to PAT events related to the driving. 
The height of the steps is non-universal and determined by the Besselfunctions. 
Since the arguments $\alpha_{L/R}$ stay constant, the step-height height decreases for large $n$ and oscillates as a function of $\alpha$.
It vanishes at nodes of the Besselfunctions, analogues to the  limit of energy-independent scattering
as studied for ac-biased junctions in Ref.~\cite{LesovikLevitov94}. 
For vanishing ac-drive only the zero order Besselfunction should contribute, 
thus we find a step height proportional to $J_0(0)=1$.
Besides the dip around $\hbar \Omega = 0$ one now expects further features in the noise at frequencies 
$\hbar \Omega=\mu_{\alpha} - \epsilon_r  \pm n \hbar \omega$ with $\alpha=L,R$.\\
\section{Current-current cross-correlations}
In this section we focus on the cross-correlation noise spectral function.
Again we write contributions to the zero temperature noise explicitly as a sum.
\begin{equation}
	S_{LR}(\Omega,\omega)=\underset{\alpha, \beta =L,R} \sum   C^{\mathrm{cross}}_{\alpha \rightarrow \beta}(\Omega,\omega) \label{SLRsum}
\end{equation} 
determines  the cross-correlation noise spectrum where the $C^{\mathrm{cross}}_{\alpha \rightarrow \beta}(\Omega)$
are in general complex quantities. If $\alpha =\beta$ the correlators $C^{\mathrm{cross}}_{\alpha \rightarrow \alpha}(\Omega) \in \mathcal{R}$ at $\hbar \Omega =0$ in the dc-limit. Accordingly, at finite frequency these terms acquire a phase factor. 
The different contributions in the dc-limit read
\begin{subequations}
 \label{SLR0cross}
\begin{align}
		C^{\mathrm{cross}}_{L \rightarrow L}(\Omega)=&  \frac{e^2 \Theta(\hbar \Omega)}{2 \pi \hbar} \underset{\mu_L-\hbar \Omega}{\overset{\mu_L}\int} d\epsilon \, \nonumber \\
&	t^*(\epsilon+\hbar\Omega) 	t(\epsilon) \left[ r^*(\epsilon)r(\epsilon+\hbar\Omega)- 1 \right]\label{correlatorscross:LtoL} \\
	%%%%%%%%%%%%%
		C^{\mathrm{cross}}_{R \rightarrow R}(\Omega)=&  \frac{e^2 \Theta(\hbar \Omega) }{2 \pi \hbar} \underset{\mu_R-\hbar \Omega}{\overset{\mu_R}\int} d\epsilon \, \nonumber \\
&
	 		t^*(\epsilon) t(\epsilon+\hbar\Omega) \left[ r^*(\epsilon+\hbar\Omega)r(\epsilon)- 1 \right]\label{correlatorscross:RtoR}\\
	%%%%%%%%%%%%%
		C^{\mathrm{cross}}_{L \rightarrow R}(\Omega)=& \frac{e^2 \Theta(\hbar \Omega-eV)}{2 \pi \hbar} \underset{\mu_R-\hbar \Omega}{\overset{\mu_L}\int} d\epsilon \,\nonumber \\
&
	 	r^*(\epsilon)t(\epsilon)  r^*(\epsilon+\hbar\Omega)t(\epsilon+\hbar\Omega)\label{correlatorscross:LtoR}\\
	%%%%%%%%%%%%%
		C^{\mathrm{cross}}_{R \rightarrow L}(\Omega)=&  \frac{e^2 \Theta(\hbar \Omega+eV)}{2 \pi \hbar}  \underset{\mu_L-\hbar \Omega}{\overset{\mu_R}\int} d\epsilon \,\nonumber \\
&
		t^*(\epsilon)r(\epsilon)t^*(\epsilon+\hbar\Omega)r(\epsilon+\hbar\Omega)\label{correlatorscross:RtoL}
\end{align}
\end{subequations}
The onsets of the $C^{\mathrm{cross}}_{\alpha \rightarrow \beta}(\Omega)$
are the same as before. As shown in Fig.~\ref{fig-6}, the finite frequency cross-correlation noise spectrum can be positive
as it is also the case in superconducting systems~\cite{Martin08,Buzdin09}. Steps in the auto-correlation spectrum now translate into peaks at negative and 
into dips at positive frequencies as it can be seen in Fig.~\ref{fig-6}. To shine a light on the this difference it is again
fruitful to study the shape of the integrands involved. In comparison to auto-correlations, 
cross-correlations exhibit a different symmetry in the pairing of s-matrices.
The cross-contributions to $S_{\alpha \beta}(\Omega,\omega)$ 
are similar to the auto-correlation contributions to $S_{\alpha \alpha}(\Omega,\omega)$ and vice versa. 
In detail, the main contribution now originates from $C^{\mathrm{cross}}_{R\rightarrow L}$. Similar to the integrand shown in Fig.~\ref{fig-2} for auto-correlation noise with ac-driving, the integrand and thus the correlator itself can be negative. 
At frequencies $\hbar \Omega \le \gamma$ the correlations between opposite terminals are 
negative by definition, due to the unitarity of the s-matrix. 
In this regime the integrand takes negative values whereas  for energies 
$\hbar \Omega \gg \gamma$ a positive contribution emerges due to PAT. 
An off-centered resonance splits the peak  at $\hbar \Omega=-eV/2$ in $C^{\mathrm{cross}}_{R\rightarrow L}$ symmetrically, in analogy to the shifting of the step position in the current-current auto-correlation spectrum, Fig~\ref{fig-1}.
If $\epsilon_r > |eV/2|$ these two peaks at  $\hbar \Omega =-eV/2 \pm \epsilon_r$ move towards $\Omega=-eV,0$ where they vanish 
and the noise spectrum gets negative along the whole emission branch ($\Omega<0$).
As for the auto-correlation noise spectrum, at $k_B \mathcal{T}=0$   the cross-correlation noise spectrum  can be calculated analytically
by assuming a Breit-Wigner lineshape, Eq.~(\ref{BreitWigner}). Integration of Eqs.(\ref{SLR0cross}) yields  
\begin{subequations} % Correlators cross - analytic @ T=0
	\label{eqn:SLR0crossResults}
\begin{align}
C^{\mathrm{cross}}_{L \rightarrow L}(\Omega)=&\Theta(\Omega)f(\Omega)(1+i\Omega/\gamma)F(\mu_L-\epsilon_r,\Omega)\label{SLRcrossResults:LtoL}\\	
C^{\mathrm{cross}}_{R \rightarrow R}(\Omega)=&\Theta(\Omega)f(\Omega)(1-i\Omega/\gamma)F(\mu_R-\epsilon_r,\Omega)\label{SLRcrossResults:RtoR}\\	
%%%%%%%%%%%%%
C^{\mathrm{cross}}_{L \rightarrow R}(\Omega)=&-\Theta(\Omega-eV)f(\Omega)\left.K(\epsilon-\epsilon_r,\Omega)\right|^{\mu_L}_{\mu_R-\Omega}\label{SLRcrossResults:LtoR}\\
C^{\mathrm{cross}}_{R \rightarrow L}(\Omega)=&-\Theta(\Omega+eV)f(\Omega)\left.K(\epsilon-\epsilon_r,\Omega)\right|^{\mu_R}_{\mu_L-\Omega}\label{SLRcrossResults:RtoL}\,,
\end{align}
\end{subequations}
where the functions $F(\epsilon,\Omega)$ and $K(\epsilon,\Omega)$ are defined in the appendix, see Eqs.~(\ref{app:def_f})-(\ref{app:def_AB}).
As for the auto-terminal noise, the correlator $C^{\mathrm{cross}}_{R \rightarrow L}(\Omega)$ is equal to  $C^{\mathrm{cross}}_{L \rightarrow R}(\Omega)$ 
when the reservoir indices $L,R$ are interchanged, and thus the voltage in the Heaviside theta function changes sign, too. 
$C^{\mathrm{cross}}_{R \rightarrow R}(\Omega)$ is equal to $C^{\mathrm{cross}}_{L \rightarrow L}(\Omega)$
if we take the complex conjugate of the pre-factor $(1+i \Omega/\gamma)$.
Overall the solutions are very similar to the auto-terminal noise spectral function were the
 most prominent difference is the imaginary part occurring in the pre-factors. Again the results can be simplified for 
a symmetric setup, leading to the replacements $\left.K(\epsilon-\epsilon_r,\Omega)\right|^{\mu_R}_{\mu_L-\Omega}\rightarrow -2 K(V/2,\Omega)$ and
$\left.K(\epsilon-\epsilon_r,\Omega)\right|^{\mu_L}_{\mu_R-\Omega}\rightarrow 2 K(V/2,-\Omega)$.
At $\Omega=0$ the noise power is given by $S_{LR}(0)=C^{\mathrm{cross}}_{R \rightarrow L}(0)$ with
\begin{align}
&C^{\mathrm{cross}}_{R \rightarrow L}(0)=	\frac{-e^2 \gamma \Theta(eV)}{4\hbar} 
\left(\frac{\gamma  (\mu_L-\epsilon_r)}{\gamma ^2+(\mu_L-\epsilon_r)^2}\right.\nonumber\\
&\left. -\frac{\gamma  (\mu_R -\epsilon_r)}{\gamma ^2+(\mu_R -\epsilon_r)^2}
+ \text{atan}\left[\frac{\mu_R -\epsilon_r}{\gamma }\right]-\text{atan}\left[\frac{\mu_L-\epsilon_r}{\gamma }\right]\right)\,
\end{align}
Assuming $\gamma \ll \mu_{L/R}-\epsilon_r$ this results in $C^{\mathrm{cross}}_{R \rightarrow L}(0)=-\frac{e^2 \gamma \pi}{2\hbar}$ 
and thus a Fano factor $F=-1/2$. At $\Omega=0$ the sum of all correlations  vanishes $S_{LR}(0)+S_{LL}(0)+S_{RL}(0)+S_{RR}(0)=0$ 
since in this limit all s-matrices are probed at the same energy.
In the  limit $|\hbar \Omega| \gg |eV|$ all correlators that contribute to the cross-correlation noise spectrum vanish. \\

Now we switch on the ac-bias voltage and set $\Omega,\epsilon_r=0$.  Then the auto-terminal contributions to the cross-correlation spectrum are real 
and can therefore be described by the product of two transmission probabilities. 
Cross-terminal contributions are related by complex conjugation. In this limit we can use the transmission functions introduced in Eq.~(\ref{transmissionFunctions}) 
to express the integrands defined by Eq.~(\ref{patnoise}) in an intuitive way as
\begin{subequations}
\begin{align}
		M^{\mathrm{cross}}_{L \rightarrow L}(\omega_{m-l},0)&= M^{\mathrm{cross}}_{R \rightarrow R}(\omega_{m-l},0) \nonumber\\
		&= T(\epsilon)T(\epsilon+(m-l)\hbar\omega)\\
		M^{\mathrm{cross}}_{L \rightarrow R}(\omega_{m-l},0)&= \left[M^{\mathrm{cross}}_{R \rightarrow L}(\omega_{m-l},0)\right]^* \nonumber	\\
		&= 	T(0,\epsilon+(m-l)\hbar\omega) R(0,\omega_{m-l} )\,.
\end{align}
\end{subequations}
We give our analytical results for the cross-correlation noise spectral function when a finite ac-bias is applied  in the appendix,  Eqs.~(\ref{app:SLRac_results}).
Corresponding noise spectra are presented in Fig.~\ref{fig-5} for different values of $a$.
\section{Energy independent scattering and elementary charge transfer processes \label{discuss}}
In the scattering approach without interaction it is straightforward to go from the 
single level setup that we have concentrated on, to two or more energy levels. 
If there is no internal coupling of the levels, the current as well as the current noise 
through the involved resonances is just the sum of the independent contributions. 
Cross-over from an energy-independent 
scattering to the multi-level case turns the straight lines shape of the noise power discussed before into a sequence of steps 
at the resonance energies. 
If the energy levels are internally coupled the difficulty is to find the corresponding 
s-matrix. For two coupled levels at zero bias voltage the frequency-dependence of shot noise 
has been studied recently~\cite{EntinWohlman07}. 
Although the resonant levels fingerprint in the spectra gives a lot of benefits when interpreting the data 
and identifying scattering channels, its energy dependence also brings a bunch of complications. 
Especially the events can not be defined by transmission and reflection probabilities, 
which connect occupied and unoccupied states in the reservoirs. 
If one drops this energy dependence, Imry et. al.~\cite{GavishImry01} have shown that
the four contributions to the noise are proportional to the Bose-distribution function $n_B(\epsilon)$.
Interestingly this originates from the product $f_{\alpha}(\epsilon)\left(1-f_{\beta}(\epsilon+\hbar\Omega) \right)$, 
which can also be written as $\left(n_B(\Omega)+1\right)\left(f_{\alpha}(\epsilon)-f_{\beta}(\epsilon+\hbar\Omega) \right)$. 
Integration over all energies yields $\hbar \Omega \left(n_B(\Omega)+1 \right)$, 
what is again proportional to the photon distribution. 
In this way the four contributions are proportional to  $M_{\beta\rightarrow \alpha} x_{\alpha \beta} (n_B(_{\alpha,\beta} )+1)$, 
with $x_{LL}=x_{RR}=\hbar \Omega$, $x_{LR}=\hbar \Omega-eV $ and $x_{RL}=\hbar \Omega+eV$.\\

In Refs.~\cite{VanevicBelzig07,VanevicBelzig08}  
the noise power has been studied for systems with time-dependent voltages 
as an interplay between {\it unidirectional} and {\it bidirectional} events of charge transfer.
Those events can be related to the four correlators of the shot-noise spectrum, 
even at energy-dependent scattering (see also Fig~\ref{fig-3}). 
Let us first set $\alpha,\hbar \Omega,k_B\mathcal{T}=0$. Then current-fluctuations are determined 
by $C_{R \rightarrow L}(\Omega,\omega)$ and are a pure source of {\it unidirectional} events. 
If there is a free state in reservoir $L$ an electron in $R$ is either reflected back to reservoir $R$ or transmitted to $L$. 
Thus, the whole process is proportional to $T(\epsilon)R(\epsilon)f^{\mathrm{e}}_{R}(\epsilon)f^{\mathrm{h}}_{R}(\epsilon)$. 
For symmetric bias $\mu_L=-\mu_R<0$ the analogues hole-like process is equivalent and describes effective electron transfer from $R$ to $L$ with 
same probability.
At finite $\Omega$ the correlator $C_{R \rightarrow L}(\Omega,0)$ 
is proportional to $T(\epsilon)R(\epsilon)f^{\mathrm{e}}_{R}(\epsilon)f^{\mathrm{h}}_{L}(\epsilon+\hbar \Omega)$. Or in terms of electron-like events
 this can be written with the help of the photonic-distribution $n_B(\epsilon)=1/[\mathrm{Exp}[\epsilon/k_B\mathcal{T} ]-1]$ as $T(\epsilon)R(\epsilon)(n_B(\Omega)+1)(f^{\mathrm{e}}_{R}(\epsilon)-f^{\mathrm{e}}_{L}(\epsilon+\hbar \Omega))$. 
Thus, we probe photonic fluctuations
due to a virtual electron-hole pair created by the frequency in lead $L$, 
with one partner being transmitted and the other one being reflected.  
$C_{L \rightarrow R}(\Omega,0)$ describes the equivalent process with electron-hole pair generation in terminal $R$ with effective charge transfer to the right. 
$C_{L \rightarrow L}(\Omega,0)$ couples electron and hole paths during reflection in the scattering region via $r^*(\epsilon)r(\epsilon+\hbar \Omega)$, what also introduces a finite scattering phase as discussed in section~\ref{sec:scattering}.
$C_{R \rightarrow R}(\Omega,0)$ then probes the difference in the transmission  of electron-hole excitations incident from the right, described by $f^{\mathrm{e}}_{R}(\epsilon)-f^{\mathrm{e}}_{R}(\epsilon+\hbar \Omega)$. 
Although auto-terminal correlators depend on a single chemical potential, rather than the bias voltage, the interplay
with PAT processes gives rise to photo-assisted \emph{unidirectional} events of charge transfer.
Now we finally examine the case of finite ac-bias $eV_{\mathrm{ac},L} \cos(\omega t)$ at $\hbar \Omega,k_B\mathcal{T}=0$.
Then both cross-contributions still describe \emph{unidirectional} ($l=0$)  and \emph{bidirectional} events ($l\ne0 $) via
\begin{align}  
    S&^{\mathrm{uni+bi}}_{LL}(\omega) =  \frac{e^2}{2\pi \hbar} \underset{l}\sum J^2_{l}(\alpha_L)\int\limits_{-\infty}^{\infty} T(\epsilon)(1-T(\epsilon))d\epsilon \nonumber\\
	 & \left( f^{\mathrm{e}}_{R}(\epsilon) f^{\mathrm{h}}_L(\epsilon-l\hbar \omega)+f^{\mathrm{e}}_L(\epsilon-l\hbar \omega)      f^h_R(\epsilon)\right)\,.\label{undirectional}
\end{align}
E.g. the first term refers to events that are proportional to $T(\epsilon)T(\epsilon_{m-l})(n_B(l\hbar\omega)+1)(f^{\mathrm{e}}_R(\epsilon)-f^{\mathrm{e}}_L(\epsilon-l\hbar \omega))$, with electron-hole pair creation in the driven ($L$) terminal for $l\ne0$. 
Auto-terminal contributions are given by 
\begin{align}  
	   S^{\mathrm{ac}}_{LL}(\omega) &= 	\frac{e^2}{2\pi \hbar}	 \underset{\beta;k,l,m}\sum    
				 J_l \left(\alpha_{\beta}\right)
				J_k\left(\alpha_{\beta}\right) 
				J_{m}\left(\alpha_{\beta}\right)
				J_{l+k-m}\left(\alpha_{\beta}\right)    \nonumber   \\
						   & \int \limits_{-\infty}^{\infty} d\epsilon   \,  T(\epsilon) T(\epsilon_{m-l})  
				  f^e_{\beta}(\epsilon_{-l}) f^h_{\beta}(\epsilon_{-k})     \, ,
\label{bidirectional}   					   
\end{align}
where $\beta=L,R$ and  $\epsilon_{n}\equiv \epsilon + n\hbar \omega$. 
Since we set $\alpha_R=0$, the term with $\beta=R$ vanishes at $k_B\mathcal{T}=0$.
This purely ac-induced contribution can not be interpreted by 
\emph{bidirectional} events. If $\beta=L$, virtual electron-hole pairs are generated in the left reservoir. 
Thus, the two particles are incident from the left, but now both species are
 transmitted with different probabilities where the whole process  is proportional to $T(\epsilon)T(\epsilon_{m-l})$.
Therefor, the correlator describes events where both particles move into the same direction. In this way both auto-terminal contributions refer to ac-induced \emph{unidirectional} charge transfer events scattered towards the measurement terminal. 
If the resonance is very narrow ($\gamma \ll eV, \hbar \omega$), the product $T(\epsilon) T(\epsilon_{m-l})$ 
will be very small if $m\ne l$. Then the main contributions from Eq.(\ref{bidirectional}) are expected when $l=m$. 
By assuming energy-independent scattering, the correlators
can be expressed in terms of the photonic distribution as 
 \begin{subequations}
 	\label{drivenwband}
\begin{align}
	 C^{\mathrm{}}_{L\rightarrow L}&=\frac{e^2 }{2\pi \hbar} T^2 \underset{k,l,m}\sum 
	 	J_l \left(\alpha_{L}\right)
	 	J_k\left(\alpha_{L}\right) 
	 	J_{m}\left(\alpha_{L}\right)
	 	J_{l+k-m}\left(\alpha_{L}\right)  \nonumber  \\
		&  \times \left[n_B((l-k)\hbar \omega)+1\right] (l-k) \hbar \omega  \\ 
%%%%%%%%%%%%%%%%%%%%%%%%
   C^{\mathrm{}}_{L\rightarrow R} &=\frac{e^2}{2\pi\hbar}  T(1-T) \underset{l}\sum J_l^2 \left(\alpha_L \right) (l\hbar \omega-eV)\nonumber \\
& \times \left[n_B (l \hbar \omega-eV)+1\right]  \\   
%%%%%%%%%%%%%%%%%%%%%%%%
   C^{\mathrm{}}_{R\rightarrow L}&= \frac{e^2  }{2\pi \hbar} T(1-T) \underset{l}\sum J_l^2 \left(\alpha_L \right) (eV-l\hbar \omega) \nonumber\\   %M^a_{R\rightarrow L}
&  \times \left[n_B (eV-l \hbar \omega)+1\right] \\
%%%%%%%%%%%%%%%%%%%%%%%%
   C^{\mathrm{}}_{R\rightarrow R}&=\frac{e^2k_B \mathcal{T} }{2\pi\hbar}  T^2 \,,
\end{align} 
 \end{subequations}
where we have assumed  $a=1$ and identical temperatures $\mathcal{T}$ in both reservoirs.
We have also dropped the arguments on the left hand side for compacter notation. On one hand, dc-induced
\emph{unidirectional events} are determined by the the cross-terminal contributions. 
On the other hand, 
 \emph{bidirectional events} are due to photonic fluctuations and 
the associated electron-hole pairs induced in the driven terminal.
This terminal ($L$) affects three out of the four correlators. If both distribution functions refer to the  ac-biased terminal, as in $C^{\mathrm{}}_{L\rightarrow L}(\omega)$, we have ac-induced \emph{unidirectional events}.\\
\section{Conclusions}
In summary we have interpreted the asymmetric noise spectra of an coherent-scattering 
double-barrier system with a single resonant level. We calculated an analytical solution 
for the photo-assisted noise spectral function for auto-terminal and cross-terminal 
current-current correlations at $k_B\mathcal{T}=0$ by assuming a Breit-Wigner lineshape for the resonance. 
At finite frequency or 
finite ac-bias shot noise is produced by partitioning of electron-hole pairs. 
As a consequence, this simple system shows a noise spectrum sensitive to many parameters. 
It exhibit signatures of quantum-coherent current-current correlations
as a sub-Poissonian Fano factor around the resonance energy. 
This anti-bunching of electrons is in competition with the PAT events, 
stimulated by the ac-driving ($n \hbar \omega$) or a static electric field  $(\hbar \Omega)$. 
At frequencies $\hbar \Omega \ge \epsilon_r + eV/2$  we find a super-Poissonian Fano factor for the auto-terminal noise 
and positive values for the cross-terminal noise when $\hbar \Omega\gg \gamma$. 
Furthermore we have shown how the scattering events can be assigned to the four different combinations 
of final and initial electronic states. Cross-terminal contributions to the auto-correlation noise spectral function
 can be related to the {\it unidirectional} and {\it bidirectional} elementary events of charge transfer identified 
in a recent microscopic derivation\cite{VanevicBelzig07}. But the scattering approach also reveals an additional kind of processes where the ac-bias voltage induces \emph{unidirectional} events directed towards the measurement terminal. In the limit $\hbar \Omega \rightarrow 0$ we could express the photo-assisted noise in terms of the photonic distribution function. The scattering 
formalism gives insight to the connection between the different regimes discussed throughout this article. 
Moreover, it also allows us to connect the interpretation of shot noise  obtained via 
different approaches, e.g. by FCS or a discussion in terms of wave packets via the \emph{Fermi golden rule}. 
The steps and dips of the noise spectra can be used in experiments to extract information about the resonance position, 
effective chemical potentials or in general to get insight into the coupling of the laser-field 
to the system in terms of PAT. 

\section{Acknowledgement}   
We would like to acknowledge financial
support by the DFG  through SFB 767 and SP 1285.
\section{Appendix: Analytic solution\label{appendix1}}
By the help of the definitions below we can write the analytic solutions for the non-symmetrized noise spectrum in a compact way.
If only dc-bias voltages are present, it turns out to be convenient to introduce the pre-factor
% Shortcuts needed only for ac-ac: prefactor
\begin{align}
	f(\Omega)=& \frac{e^2 \gamma^3}{2\left(4\gamma^2+\Omega^2 \right)}	\label{app:def_f}\,.
\end{align}
Furthermore we use the expressions
% Shortcuts needed only for ac-ac
\begin{align}
	F(\epsilon,\Omega)&= \text{atan}\left(\frac{\epsilon+\Omega}{\gamma} \right) - \text{atan}\left(\frac{\epsilon-\Omega}{\gamma} \right)	\nonumber \\
	-&\frac{\gamma}{\Omega} \ln\left[\frac{ (\gamma^2+\epsilon^2)^2}{\left( \gamma^2+\left(\epsilon+\Omega\right)^2\right)\left( \gamma^2+\left(\epsilon-\Omega\right)^2\right)}\right] \label{app:def_F}
\end{align}
\begin{align}
	G(\epsilon,\Omega)&= \left[3+\left(\frac{\Omega}{\gamma}\right)^2\right]\text{atan}\left(\frac{\epsilon+\Omega}{\gamma} \right) - \text{atan}\left(\frac{\epsilon}{\gamma} \right)	\nonumber \\
	+&\frac{\gamma}{\Omega} \ln\left[\frac{ \gamma^2+\epsilon^2}{ \gamma^2+\left(\epsilon+\Omega\right)^2}\right]
\end{align}
\begin{align}
	H(\epsilon,\Omega)&= \left[2+\left(\frac{\Omega}{\gamma}\right)^2\right]\left(\text{atan}\left(\frac{\epsilon+\Omega}{\gamma} \right) + \text{atan}\left(\frac{\epsilon}{\gamma} \right)	\right)\nonumber \\
	+&2\frac{\gamma}{\Omega} \ln\left[\frac{ \gamma^2+\epsilon^2}{ \gamma^2+\left(\epsilon+\Omega\right)^2}\right]
	\label{app:def_GH}
\end{align}
for auto- and cross-terminal noise. To achieve a compact notation for the  cross-terminal noise we also need the definition
% Shortcuts needed only for dc-cc
\begin{align}
	K(\epsilon,\Omega)&= \text{atan}\left(\frac{\epsilon+\Omega}{\gamma} \right) + \text{atan}\left(\frac{\epsilon}{\gamma} \right)	\nonumber \\
	+&\frac{\gamma^2+\Omega^2/2}{\gamma   \Omega} \ln\left[\frac{ \gamma^2+\epsilon^2}{ \gamma^2+\left(\epsilon+\Omega\right)^2}\right]\label{app:def_K}\,.
\end{align}
% Shortcuts needed only for ac-ac + ac-cc: pre-factor
If additional ac-bias voltages are present it is reasonable to make use of the pre-factor
\begin{align}
	\tilde{f}(\Omega)=\frac{e^2 \gamma^4}{4} \label{app:def_ftilde}
\end{align}
% Shortcuts needed only for ac-ac + ac-cc
and the shorthands
\begin{align}
	D_1=&\left[(2 i \gamma +\Omega)(2 i \gamma + \omega)(\Omega + \omega)\right]^{-1}\\
	D_2=&\left[(2 i \gamma +\Omega)(-2 i \gamma + \omega)(\Omega - \omega)\right]^{-1}\\
% \end{align}
% % Shortcuts needed only for ac-cc
% \begin{align}
	D^{\pm}_3=& 2 \gamma(i\gamma \pm\Omega+\omega)\\
	D^{\pm}_4=& 2 (\gamma\pm i\Omega)(i\gamma+\omega)\,. \label{app:def_D}
\end{align}
Finally we complete the set of functions with
% Shortcuts needed only for ac-ac + ac-cc
\begin{align}
	A^{\pm}(\epsilon,\Omega,\omega)&=2i \text{atan}\left(\frac{\epsilon+\Omega+\omega}{\gamma}\right)\nonumber \\
	 \pm &\ln\left(\gamma^2+(\epsilon+\Omega+\omega)^2\right)\\
	B^{\pm}(\Omega,\omega)&= 2 (\gamma+i\Omega)(\pm i \gamma +\Omega+\omega) \label{app:def_AB}\,,
\end{align}
where $A^{\pm}(\epsilon,\Omega,\omega)$ defines the basic shape of the results for ac-biased systems and $B^{\pm}(\Omega,\omega)$ is needed
for the description of the cross-correlation spectrum.
Below we present the results for the photo-assisted noise spectral density of auto-terminal and cross-terminal current-current correlations.
We assume a Breit-Wigner lineshape (\ref{BreitWigner}) for the resonant level and perform the energy integration in Eqn.(\ref{patnoise}). The results are plotted as a function of frequency in Fig.\ref{fig-5}. Due to the cumbersome expressions we 
use the shorthands defined above as well as the notation $\tilde{\omega}\equiv (m-l)\hbar \omega$ 
and set $\hbar=1$. For the auto-correlation function we then find
\begin{widetext}
% Driven setup: auto-correlations %%%%%%%%%%%%%%%%%%%%%%%%%
%
\begin{subequations}\label{app:SLLac_results}
\begin{align}
% CLtoL ac
	C^{\mathrm{}}&_{L \rightarrow L}(\Omega,\omega)=\tilde{f}(\Omega)(1+(\Omega/\gamma)^2)\sum\limits_{lkm} \Theta( \Omega +(l-k)\omega)
	J_l\left(\alpha_L \right) J_k\left(\alpha_L \right) J_{m+k-l}\left(\alpha_L  \right) J_m\left(\alpha_L  \right)\nonumber\big[	\big.\nonumber \\
	& \big.-D_1A^-(\epsilon,0,0)+D_2 A^-(\epsilon,\Omega,0)-D_2^*A^-(\epsilon,\Omega,\tilde{\omega})+D_1^*A^-(\epsilon,0,\tilde{\omega})\big]^{\epsilon=\mu_L+l \omega}_{\epsilon=\mu_R- \Omega + k \omega}\\
% CRtoR ac
			C^{\mathrm{}}&_{R \rightarrow R}(\Omega,\omega)=\tilde{f}(\Omega)\sum\limits_{lkm} \Theta( \Omega +(l-k)\omega)
		J_l\left(\alpha_L \right) J_k\left(\alpha_L \right) J_{m+k-l}\left(\alpha_L  \right) J_m\left(\alpha_L  \right)\nonumber	\big[	\big.\nonumber \\
			& \big.D_1A^+(\epsilon,0,0)+D_1^*A^-(\epsilon,0,\tilde{\omega})-D_2^*A^-(\epsilon,\Omega,0)-D_2^*A^+(\epsilon,\Omega,\tilde{\omega})
		\big]^{\epsilon=\mu_R+l \omega}_{\epsilon=\mu_R- \Omega + k  \omega}\\
% CLtoR ac
		C^{\mathrm{}}&_{L \rightarrow R}(\Omega,\omega)= \tilde{f}(\Omega)\sum\limits_{lkm}\Theta( \Omega+(l-k)\omega-eV ) 	J_l\left(\alpha_L \right) J_k\left(\alpha_R \right) J_{m+k-l}\left(\alpha_R  \right) J_m\left(\alpha_L  \right)\nonumber\big[	\big.\nonumber \\
			& \big. B^+(0,\tilde{\omega})D_1 A^+(\epsilon,0,0)+ B^-(\Omega,\tilde{\omega})D_1^*A^+(\epsilon,\Omega,\tilde{\omega})+ B^-(-\Omega,\tilde{\omega})D_2A^+(\epsilon,\Omega,0)+ B^+(0,\tilde{\omega})D_2^*A^-(\epsilon,0,\Omega)
		\big]^{\epsilon=\mu_L+l \omega}_{\epsilon=\mu_R- \Omega + k  \omega}\\
% CRtoL ac
			C^{\mathrm{}}&_{R \rightarrow L}(\Omega,\omega)= \tilde{f}(\Omega)\sum\limits_{lkm} \Theta( \Omega+(l-k)\omega+eV )J_l\left(\alpha_R \right) J_k\left(\alpha_L \right) J_{m+k-l}\left(\alpha_L  \right) J_m\left(\alpha_R  \right)\nonumber\big[	B^+(0,\tilde{\omega})D_1A^-(\epsilon,0,0)+\big.\nonumber \\
			& \big.+ B^-(-\Omega,\tilde{\omega})D_2A^-(\epsilon,\Omega,0)+ B^+(0,\tilde{\omega})D_2^*A^+(\epsilon,0,\Omega)+ B^-(\Omega,\tilde{\omega})D_1^*A^+(\epsilon,\Omega,\tilde{\omega})
		\big]^{\epsilon=\mu_L+l  \omega}_{\epsilon=\mu_L- \Omega + k  \omega} \,.
\end{align}
\end{subequations}
%
%
%
%
%
%
% Driven setup: cross-correlations % (ac-cc) %%%%%%%%%%%%%%%%%%%%%%%
%
%
%
%
%
Using the same notation the solution of the cross-terminal correlations can be cast in the form
\begin{subequations}\label{app:SLRac_results}
\begin{align}
	% CcLtoL ac
			C^{\mathrm{cross}}_{L \rightarrow L}(\Omega, \omega)&=\sum\limits_{lkm} \Theta( \Omega +(l-k)\omega)
		J_l\left(\alpha_L \right) J_k\left(\alpha_L \right) J_{m+k-l}\left(\alpha_L  \right) J_m\left(\alpha_L  \right) \big[ \big.\nonumber\\
		& \big.  D_1A^+(\epsilon,0,0)-D_1^*A^+(\epsilon,\Omega,\tilde{\omega})-D_2A^+(\epsilon,\Omega,0)
				+D_2^*A^-(\epsilon,0,\tilde{\omega})
			\big]^{\epsilon=\mu_L + l \omega}_{\epsilon=\mu_L - \Omega+ k \omega}\\
	% CcRtoR ac
			C^{\mathrm{cross}}_{R \rightarrow R}(\Omega, \omega)&=  \sum\limits_{lkm} \Theta( \Omega +(l-k)\omega)
		J_l\left(\alpha_R \right) J_k\left(\alpha_R \right) J_{m+k-l}\left(\alpha_R  \right) J_m\left(\alpha_R  \right) \big[ \big.\nonumber\\
		& \big.  D_1A^+(\epsilon,0,0)-D_1^*A^-(\epsilon,\Omega,\tilde{\omega})-D_2A^-(\epsilon,\Omega,0)
				+D_2^*A^+(\epsilon,0,\tilde{\omega})
			\big]^{\epsilon=\mu_R + l \omega}_{\epsilon=\mu_R - \Omega+ k \omega}\\
	% CcLtoR ac
			C^{\mathrm{cross}}_{L \rightarrow R}(\Omega, \omega)&= \sum\limits_{lkm}\Theta( \Omega+(l-k)\omega-eV ) J_l\left(\alpha_L \right) J_k\left(\alpha_R \right) J_{m+k-l}\left(\alpha_R  \right) J_m\left(\alpha_L  \right) \big[ \big. \nonumber\\
		& \big. D^+_3 D_1A^+(\epsilon,0,0)+(D_3^+)^*D_1^*A^-(\epsilon,\Omega,\tilde{\omega})+(D_4^+)^* D_2A^+(\epsilon,\Omega,0)
				+D_4^+ D_2^*A^-(\epsilon,0,\tilde{\omega})
			\big]^{\epsilon=\mu_L + l \omega}_{\epsilon=\mu_R - \Omega+ k \omega}\\
	% CcRtoL ac
			C^{\mathrm{cross}}_{R \rightarrow L}(\Omega, \omega)&=\sum\limits_{lkm}\Theta( \Omega+(l-k)\omega+eV )J_l\left(\alpha_R \right) J_k\left(\alpha_L \right) J_{m+k-l}\left(\alpha_L  \right) J_m\left(\alpha_R  \right)\big[ \big.\nonumber\\
			& \big. D^-_4 D_1A^-(\epsilon,0,0)+(D_4^-)^*D_1^*A^+(\epsilon,\Omega,\tilde{\omega})+(D_3^-)^* D_2A^-(\epsilon,\Omega,0)
					+D_3^-  D_2^*A^+(\epsilon,0,\tilde{\omega})
			\big]^{\epsilon=\mu_R + l \omega}_{\epsilon=\mu_L - \Omega+ k \omega}\,,
\end{align}
\end{subequations}
 In the dc-limit these expressions simplify to Eqs.~(\ref{eqn:SLL0autoResults}) for auto-correlation noise and Eqs.~(\ref{eqn:SLR0crossResults}) for cross-correlation noise. Obviously, the additional ac-bias introduces a complicated 
 scattering phase via the imaginary parts in the above expressions. The noise spectrum is plotted for different asymmetry parameters $a$ in Fig.~\ref{fig-5}.
Ac-bias voltages introduce additional peaks and dips related to the driving frequency $\omega$. By varying $a$, such PAT induced peaks in the cross-correlation noise spectra can turn into dips and vice versa.
\end{widetext}
\section*{References}

\end{document}